\documentclass[a4paper,12pt,man,british]{apa6}
\usepackage[hang,flushmargin]{footmisc} 
\usepackage[american]{babel}
\usepackage[utf8x]{inputenc}
\usepackage{amsmath}
\usepackage{graphicx}
\usepackage{epstopdf}
\usepackage{apacite}
\usepackage{array}
\usepackage{rotating}
\usepackage{threeparttablex}
\usepackage{environ}
\usepackage{hyphenat}
\usepackage{setspace}
\usepackage{footmisc}

\title{The Social Benefits of Balancing Creativity and Imitation:
Evidence from an Agent-based Model}
\shorttitle{Social Benefits Balancing Creativity Imitation}
\author{Liane Gabora and Simon Tseng}
\affiliation{University of British Columbia}

\abstract{Although creativity is encouraged in the abstract it is often discouraged in educational and workplace settings. Using an agent-based model of cultural evolution, we investigated the idea that tempering the novelty-generating effects of creativity with the novelty-preserving effects of imitation is beneficial for society. 
In Experiment One we systematically introduced individual differences in creativity, and observed a trade-off between the ratio of creators to imitators, and how creative the creators were. Excess creativity was detrimental because creators invested in unproven ideas at the expense of propagating proven ones.
Experiment Two tested the hypothesis that society as a whole benefits if individuals adjust how creative they are in accordance with their creative success. When effective creators created more, and ineffective creators created less (social regulation), the agents segregated into creators and imitators, and the mean fitness of outputs was temporarily higher. We hypothesized that the temporary nature of the effect was due to a ceiling on output fitness. 
In Experiment Three we made the space of possible outputs open-ended by giving agents the capacity to chain simple outputs into arbitrarily complex ones such that fitter outputs were always possible. With the capacity for chained outputs, the effect of social regulation could indeed be maintained indefinitely. The results are discussed in light of empirical data.\\
\vspace{12pt}
}

\authornote{
Accepted for publication in {\it Psychology of Aesthetics, Creativity, and the Arts}. This article may not exactly replicate the final version published in the APA journal. It is not the copy of record.

Correspondence concerning this manuscript may be addressed to: Liane Gabora, Department of Psychology, University of British Columbia, Fipke Centre for Innovative Research, 3247 University Way, Kelowna, British Columbia, Canada, V1V 1V7; Email: liane.gabora@ubc.ca

This work was supported by a grant (62R06523) from the Natural Sciences and Engineering Research Council of Canada. Thanks to Fabian Y\'{a}\~{n}ez for assistance with the manuscript, and to Tiha von Ghyczy for assistance with the analysis. Preliminary reports on portions of this work were presented at the Annual Meeting of the Cognitive Science Society and the International Conference on Computational Creativity.}

\begin{document}
\maketitle
\section{Introduction}
\label{sec:intro}
Creativity is praised as the hallmark of our humanity, responsible for our greatest achievements (Mithen, 1998). It is essential for maintaining a competitive edge in the marketplace (Reiter-Palmon \& Illies, 2004; Rule \& Irwin, 1988), and has long been associated with personal fulfillment (May, 1975; Rogers, 1959), self-actualization (Maslow, 1959), and more recently with the positive psychology movement (Adams, 2012; Kaufman \& Beghetto, 2014; Simonton, 2002).
However, 
social norms, policies, and institutions often stifle creativity (Ludwig, 1995; Sulloway, 1996), and educational systems do not appear to prioritize the cultivation of creativity, and in some ways discourage it (Snyder, Gregerson, \& Kaufman, 2012; Robinson, 2001). 
Teachers often have conscious or unconscious biases against creative students, 
leading them to act in ways that suppress creativity (Aljughaiman \& Mowrer-Reynolds, 2005; Beghetto, 2007; Plucker, Beghetto, \& Dow, 2004; Westby \& Dawson, 1995). Workplaces often discourage creativity by providing insufficient resources and support for the development of new ideas, and inappropriate levels of challenge and autonomy (Amabile, 1998), as well as levels of environmental distraction that are not conducive to creativity (Stokols, Clitheroe, \& Zmuidzinaz, 2002). 
Is there any rhyme or reason to society's mixed messages about the desirability of creativity?

\subsection{Balancing Novelty with Continuity}
There are drawbacks to creativity (Cropley, Cropley, Kaufman, \& Runco, 2010; Ludwig, 1995), one being that generating creative ideas is difficult and time consuming. Moreover, a creative solution to one problem often generates other problems, or has unexpected negative side effects that only become apparent after much effort has been invested (Tomlinson, 1980).
Given the costs of creativity, 
it seems reasonable to speculate that there may be an adaptive value to the seemingly mixed messages that society sends about the desirability of creativity; perhaps society is well-served by the tension between creative expression and the reinforcement of conventions and established protocols. 

This paper explores the possibility that mechanisms at work encouraging individual differences in creativity could be beneficial, by ensuring that the society as a whole both generates new variants and preserves the best of them.
This would be consistent with growing evidence that group behaviour does not always reduce to individual behaviour (e.g., Anderson, Richardson, \& Chemero, 2012; Goldstone \& Gureckis, 2009). It is also consistent with our everyday experience that an extended social group can reap the rewards of the creative efforts of an individual, i.e., few of us would be able to build a computer or write a symphony, but they are nonetheless ours to use and enjoy. We all benefit from the exchange of knowledge, ideas, and artifacts, in part because of our capacity for \emph{social learning}, a phenomenon that Bandura (1995) described as `no-trial learning', which involves learning by observing and imitating others.

In much of the cultural evolution literature, social learning is contrasted with individual learning, which involves learning for oneself, and novelty is attributed to things like copying error (e.g., Henrich \& Boyd, 2002; Mesoudi, Whiten \& Laland, 2006; Rogers, 1988). Creativity, if mentioned at all, is equated with individual learning. However, they are not the same thing. Individual learning deals with obtaining pre-existing information from the environment through non-social means (e.g., reading a book), whereas creativity involves generating ideas, behavior, or artifacts that did not previously exist. In the first case the information comes from the external world; in the second it is generated internally. Indeed there is increasing recognition of the extent to which creative outcomes are contingent upon internally driven incremental/iterative processing (Basadur, 1995; Chan \& Schunn, 2015; Feinstein, 2006). 

It is well known in theoretical biology that  cumulative evolution entails a fusion of \emph{variation generating} processes, such as mutation, and processes that \emph{preserve fit variants}, such as heredity (Haldane, 1932). It has been suggested that in cultural evolution the role of variation generation is played by creativity, and the role of variation preservation is played by social learning processes such as imitation (Gabora, 1995). 
``In vivo'' studies of scientific laboratories reveal that scientists benefit from opportunities for distributed reasoning and scaffolding of ideas and interpretations afforded by social networks (Dunbar, 2000). Similarly, through the interplay of creativity and social learning, ideas in the arts, sciences, and technology, as well as customs and folk knowledge, exhibit recombinant growth (Weitzman, 1998), and evolve over time (Dasgupta, 1994; Jacobs, 2000). The pattern of cumulative cultural change that results when new innovations draw from and build upon on existing products is sometimes referred to as the \emph{ratchet effect} (Tomasello, Kruger, \& Ratner, 1993). 

There is also evidence (reviewed in Hills, Todd, Lazer, Redish, \& Couzin, 2015) that firms as well as societies benefit by balancing exploration with exploitation. The finding of successful solutions is made possible through exploration, while social learning processes such as imitation assist in the perpetuation and exploitation of these successful solutions, and continuity is provided by the maintenance and diffusion of routines, which must evolve in response to changing markets. 
Organizational leaders need to provide employee autonomy and be on the lookout for opportunities emerging from employee efforts, yet balance this with the provision of sufficient constraints to make goals seem within reach, and the pruning out of inferior ideas (Hunter, Thoroughgood, Myer, \& Ligon, 2011; Mumford \& Hunter, 2005). 
Further evidence for the notion that productivity involves a balancing of novelty and continuity comes from a study of alliances between firms based on data for 116 companies in the chemicals, automotive and pharmaceutical industries (Nooteboom, Van Haverbeke, Duysters, Gilsing, \& Van den Oord, 2007). The authors found an inverted U-shaped effect of cognitive distance on innovation, where cognitive distance was operationalized in terms of differences in technological knowledge between the two firms, and innovativeness was assessed through an analysis of patent applications.
They concluded that alliances between firms with low cognitive distance introduces too little novelty to increase productivity, while high cognitive distance means insufficient continuity for cumulative knowledge growth. 

In short, it seems reasonable that the mixed messages society gives about the desirability of creativity might stem from society's need to balance novelty generation with novelty preservation, which can be understood in terms of theoretical considerations of culture as an evolution process. 

\subsection{Agent-based Models}
This interplay between `exploration / generation of novelty' and `exploitation / perpetuation of novelty' can be examined with an agent-based model.
An agent-based model (ABM) is a computer program that simulates the actions and interactions of autonomous agents (both individual or collective entities such as organizations or social groups) in order to assess their effects on the system as a whole (for a review of ABMs see Niazi \& Hussain, 2011). 
Because ABMs enable us to manipulate variables and observe the effects in a more controlled manner than in real life, they have proven useful for investigating questions concerning the diffusion of creative novelty and its impact on cultural evolution (e.g., Gabora, 2008a, 2008b; Guardiola, Diaz-Guilera, Perez, Arenas, \& Llas, 2002; Iribarren \& Moro, 2011; Jackson \& Yariv, 2005; Liu, Madhavan, \& Sudharshan, 2005; Sosa \& Connor, 2015; Spencer, 2012; Watts \& Gilbert, 2014). For example, results obtained with ABMs suggest that agents in large, diverse populations tend to be more creative (Gabora, 2008a; Spencer, 2012), the density of communication links amongst agents produces diminishing returns in term of the benefits on the invention rate (Bhattacharyya \& Ohlsson, 2010), and diverse communities are better at generating novelty while communities of specialized agents may be better at communicating novelty Spencer, 2012). 

Some computational models referred to as models of cultural evolution (e.g., Henrich \& Boyd, 2002) allow for as few as only two alternative forms of a cultural trait, i.e., there is no accumulative ratcheting of novelty. They are thus properly referred to as models of cultural transmission, not models of cultural evolution. 
However, others do allow for genuine accumulation of novelty. In MAV (for `meme and variations'), an ABM of cultural evolution (Gabora, 1995), and precedessor of the model used here, novelty was injected into the artificial society through the invention of new actions, and continuity was preserved through the imitation of existing actions. When agents never invented, there was nothing to imitate, and there was no cultural evolution. 
Indeed, it makes intuitive sense that if everyone relies on the strategy of copying others rather than coming up with their own ideas, there are no new ideas around to imitate, and the generation of cultural novelty grinds to a halt. 
If the ratio of invention to imitation in MAV was even marginally greater than 0, not only was cumulative cultural evolution possible, but eventually all agents converged on optimal outputs. When all agents always invented and never imitated, the mean fitness of cultural outputs was also sub-optimal because fit ideas were not dispersing through society. (In this cultural context, \emph{fitness} refers to value for the agent according to a fitness function, as discussed at length below.) The society as a whole performed optimally when the ratio of creating to imitating was approximately 2:1. \footnote{Note that this finding cannot be construed as support for Rogers' (1988) claim that cheap social learning does not necessarily increase mean fitness, for several reasons, one being that in MAV, and in the current work, the concern is the fitness of cultural outputs, not the biological fitness of individuals. These are sometimes related, but not necessarily, and indeed sometimes at odds with one another (as when tasks such as preparing food and caring for offspring are neglected due to immersion in a creative project).}
Extreme levels of creativity were detrimental at the level of the society, suggesting that there could be an adaptive value to society's ambivalent attitudes toward creativity.

\subsection{Hypotheses and Approach}
This paper provides a computational test of three hypotheses that have not previously been explored in the ABM literature, hypotheses that challenge the common assumption that more creativity is necessarily better. First, we tested the hypothesis that society as a whole can suffer if either (1) the ratio of creators to imitators is too high, or (2) creators are too creative. Although experiments with MAV had shown that the mean fitness of cultural outputs decreased if agents were too creative, in those experiment all agents were equally creative. Findings of pronounced individual differences in creativity (Kaufman, 2003; Wolfradt \& Pretz, 2001) suggested that a logical next step was to investigate how varying the extent of such individual differences impacts the society as a whole. 

The second hypothesis tested here is that a society can perform better if individuals are able to adjust how creative they are over time in accordance with their perceived creative success. There is empirical evidence that children can adjust their imitative fidelity and level of innovation (Legare, Wen, Herrmann, \& Whitehouse, 2015), and that high imitative fidelity can be related to fear of ostracism (Watson-Jones, Legare, Whitehouse, \& Clegg, 2014). Thus, society may balance novelty and continuity through mechanisms such as selective ostracization of deviant behaviour unless it is accompanied by the generation of valuable creative output, and encouragement or even adulation of those whose creations are successful. In this way society might self-organize into a balanced mix of novelty generating creators and continuity perpetuating imitators, both of which are necessary for cumulative cultural evolution. In theory, if effective creators create more, and ineffective creators create less, the society's outputs should collectively evolve faster. 

A first step in investigating this was to determine whether it is algorithmically possible to increase the mean fitness of ideas in a society by enabling agents to self-regulate how creative they are. We refer to this regulatory mechanism as \emph{social regulation} (SR) because it could be mediated by social cues such as praise and/or criticism from peers, family, or teachers, but it is also possible that it involves individual differences in the ability to detect or respond to such cues, or individuals' own assessments of the worth of their ideas, or some combination of these.

A third hypothesis investigated here is that in order for the benefit of this social regulation mechanism to be ongoing (as opposed to temporary), the space of possible creative outputs must be open-ended, such that it is always possible for superior possibilities to be found. In other words, social regulation is advantageous only when it is possible to obtain fitter outputs than those currently in use. 

\section{The Computational Model}
The ABM used here, referred to as ``EVOlution of Culture'', abbreviated EVOC, is a model of cultural evolution that uses neural network based agents that (1) invent new ideas, (2) imitate actions implemented by neighbors, (3) evaluate ideas, and (4) implement successful ideas as actions (Gabora, 2008a).\footnote{The code is freely available; to gain access please contact the first author.} EVOC was used because it is amenable to testing the above hypotheses concerning creativity; discussion of general questions about how culture evolves including comparison with other approaches (e.g., Boyd \& Richerson, 1985) can be found elsewhere (Gabora, 2008b, 2011, 2013; Gabora \& Kauffman, 2016). The approach is consistent with a growing effort in cognitive science to leverage computer modeling techniques and knowledge of cognition to understand aggregate social outcomes (Goldstone \& Gureckis, 2009). 

EVOC is an elaboration of the above-mentioned MAV (Gabora, 1995), the earliest computer program to isolate culture as an evolutionary process in its own right so that it can be compared and contrasted with biological evolution.\footnote{The approach can thus be contrasted with computer models of cultural transmission, in which (unlike models of cultural evolution) there may be as few as two possible outputs, and the outputs do not become increasingly complex and adapted over time, and with computer models of how individual learning affects biological evolution (Best, 1999; Higgs, 2000; Hinton \& Nowlan, 1987; Hutchins \& Hazelhurst, 1991).} The goal behind MAV, and also behind EVOC, was to distill the underlying logic of cultural evolution, i.e., the process by which ideas adapt and build on one another in the minds of interacting individuals. Agents do not evolve in a biological sense, as they neither die nor have offspring, but do in a cultural sense, by generating and sharing ideas for actions. The cultural outputs in EVOC take the form of actions, since Donald (1991) and others have provided substantial evidence that the earliest elements to evolve through culture, before grammatical language, were physical actions such as gestures, and the movements required to make tools. 

EVOC has been used to address such questions as how does the presence of leaders or barriers to the diffusion of ideas affect the fitness and diversity of cultural outputs (Gabora, 2008b). Here, we use it to investigate the social impact of varying the ratio of creators to imitators and enabling social regulation of individual creativity levels.

We now summarize the architecture of EVOC in sufficient detail to explain our results.

\subsection{Agents}
Agents consist of (1) a neural network, which encodes ideas for actions and detects trends in what constitutes a fit action, (2) a ``perceptual system'', which observes and evaluates neighbors' actions, and (3) a body, consisting of six body parts which implement actions.

\begin{figure}
\centering
\includegraphics[width=0.95\columnwidth]{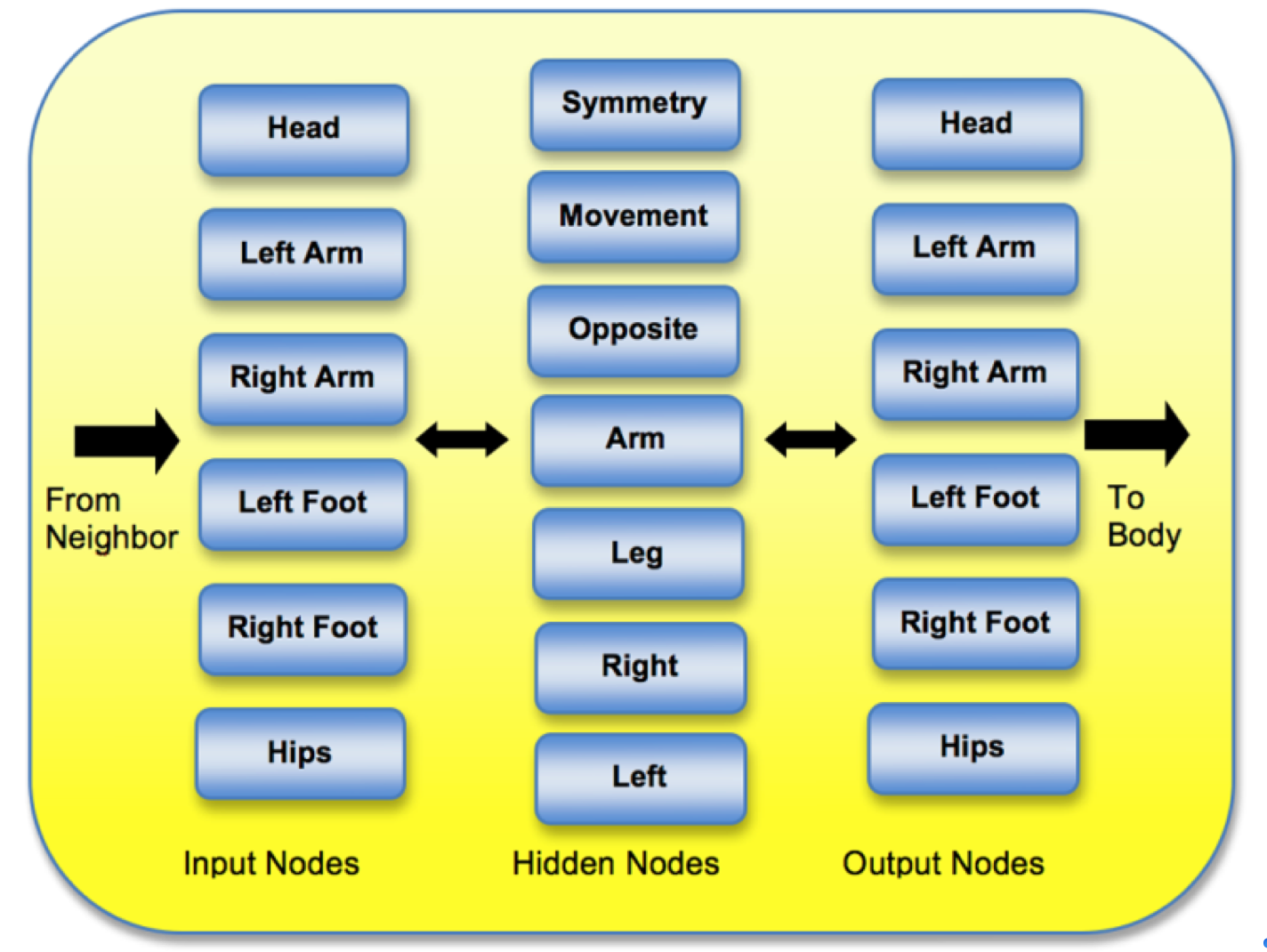}
\caption{The core of an agent is an auto-associative neural network composed of six input nodes and six corresponding output nodes that represent concepts of body parts (LEFT ARM, RIGHT ARM, LEFT LEG, RIGHT LEG, HEAD, and HIPS), and seven hidden nodes that represent more abstract concepts (LEFT, RIGHT, ARM, LEG, SYMMETRY, MOVEMENT and OPPOSITE). Input nodes and output nodes are connected to hidden nodes of which they are instances (e.g. RIGHT ARM is connected to RIGHT.) The hidden nodes are used to bias invention using learned trends about what constitutes a fit action.}
\label{fig:network}
\end{figure}

The neural network is an auto-associator because this enables the agent to learn and execute the action that a neighbor is executing, and thereby imitate successful neighbors.\footnote{Learning in auto-associative networks is unsupervised in the sense that they take in inputs, and try to organize internal representations based on them.} The network is composed of six input nodes and six corresponding output nodes that represent concepts of body parts (LEFT ARM, RIGHT ARM, LEFT LEG, RIGHT LEG, HEAD, and HIPS), and seven hidden nodes that represent more abstract concepts (LEFT, RIGHT, ARM, LEG, SYMMETRY, OPPOSITE, and MOVEMENT). Input nodes and output nodes are connected to hidden nodes of which they are instances (e.g., RIGHT ARM is connected to RIGHT). A schematic illustration of the neural network is provided in Figure \ref{fig:network}. Each body part can occupy one of three possible positions: a neutral or default position, and two other positions, which are referred to as active positions. Activation of any input node activates the MOVEMENT hidden node. Same-direction activation of symmetrical input nodes (e.g., positive activation---which represents upward motion---of both arms) activates the SYMMETRY node. 

\begin{center} 
Insert Figure \ref{fig:network} here.
\end{center}

The neural network starts with small random weights between input/output nodes. Weights between hidden nodes, and weights between hidden nodes and input/output nodes, are fixed at +/- 1.0. Patterns that represent ideas for actions are learned by training for 50 iterations using the generalized delta rule with a sigmoid activation function (Rumelhart \& McClelland, 1986). (See the Appendix for details.) Training continues until it has learned the identity function between input and output patterns. 

The neural network enables agents to learn trends over time concerning what general types of actions tend to be valuable (e.g., that symmetrical actions tend to be fit), and use this learning to invent new actions more effectively (e.g., to increase the frequency of symmetrical actions). When the ability to learn such trends is turned off, agents invent at random and the fitness of their inventions increases much more slowly (Gabora, 2008b). 

\subsection{Invention}
An idea for a new action is a pattern consisting of six elements that dictate the placement of the six body parts. Agents generate new actions by modifying their initial action or an action that has been invented previously or acquired through imitation. During invention, the pattern of activation on the output nodes is fed back to the input nodes, and invention is biased according to the activations of the SYMMETRY and MOVEMENT hidden nodes. (Note that, were this not the case, there would be little point in using a neural network. Note also that while in the first iteration the agent is simply guessing and learning, over the course of a run, invention becomes increasingly more sophisticated.) To invent a new idea, for each node of the idea currently represented on the input layer of the neural network, the agent makes a probabilistic decision as to whether the position of that body part will change, and if it does, the direction of change is stochastically biased according to the learning rate. If the new idea has a higher fitness than the currently implemented idea, the agent learns and implements the action specified by that idea. When ``chaining'' is turned on (as discussed below), an agent can keep adding new sub-actions and thereby execute a multi-step action, so long as the most recently-added sub-action is both an acceptable sub-action and different from the previous sub-action of that action (Gabora, Chia, \& Firouzi, 2013).

\subsection{Imitation}
The process of finding a neighbor to imitate works through a form of lazy (non-greedy) search. The imitating agent randomly scans its neighbors, and adopts the first action that is fitter than the action it is currently implementing. If it does not find a neighbor that is executing a fitter action than its own current action, it continues to execute the current action.

\subsection{Evaluation: The Fitness Function}
Following (Holland, 1975), we refer to the success of an action in the artificial world as its fitness, with the caveat that unlike its usage in biology, here the term is unrelated to number of offspring (or number of ideas derived from a given idea). As mentioned previously, the fitness function in EVOC involves bodily movement, on the basis of evidence that the earliest elements to evolve through culture were physical actions. The fitness function used in the first two experiments rewards activity of all body parts except for the head, symmetrical limb movement, and positive limb movement. The rationale for this is that many human actions require a stationary head (to watch what you're doing), and symmetrical limb movement, i.e., these are relatively common constraints on many real movements.\footnote{Of course, these constraints are not present for {\it all} human activities, such as holding a yoga posture.}
The fitness function was also designed to meet practical constraints, such as having multiple optima (e.g., an action can be optimal if either both arms move up or both arms move down.) Multiple optima enables us to better characterize the effect of a given manipulation on diversity (i.e., whether it finds all optima or just one).\footnote{Another reason this fitness function was used is that it exhibits a cultural analog of epistasis which makes it more difficult to solve. In biological epistasis, the fitness conferred by the allele at one gene depends on which allele is present at another gene. In this cognitive context, epistasis is present when the fitness contribution to the idea by movement of one limb depends on what other limbs are doing.}

Total body movement, $m$, is calculated by adding the number of active body parts, \emph{i.e.,} body parts not in the neutral position. $m_u$ is the number of body parts moving upwards.\\

\vspace{2ex}
\hspace*{\fill}%
\begin{tabular}{l@{}p{2in}p{2in}}
$m_h = 1$ & if head is stationary;      &  0 otherwise \\
$s_a = 1$ & if arms move symmetrically; & 0 otherwise \\
$s_l = 1$ & if legs move symmetrically; & 0 otherwise \\
$p_a = 1$ & if arms move upwards;       & 0 if arms move downwards \\
$p_l = 1$ & if legs move upwards;       & 0 if legs move downwards \\
\end{tabular}%
\hspace*{\fill}
\vspace{2ex}

\noindent Fitness of a single-step action, $F_n$, is determined as follows: 
\begin{equation}
F_{n} = m + 2m_u + 10m_h + 5(s_a+s_l) + 2(p_a+p_l)  
\label{eq:fitnesssingle}
\end{equation}

The weights reflect intuitive notions about the relative importance of different aspects of what makes for a fit action. For example, since (as mentioned previously) almost all actions require that the head remain stationary so as to be able to focus on stimuli of interest, the weight on $m_h$ is very high, and since (as also mentioned previously) many actions require symmetrical movement, the weight on $s_a$ and $s_l$ are moderately high.

\subsection{Learning}
Invention makes use of the ability to detect, learn, and respond adaptively to trends. Since no action acquired through imitation or invention is implemented unless it is fitter than the current action, new actions provide valuable information about what constitutes an effective idea. Knowledge acquired through the evaluation of actions is translated into educated guesses about what constitutes a successful action using weight updating through feedback. For example, an agent may learn that more overall movement tends to be either beneficial (as with the fitness function used here) or detrimental, or that symmetrical movement tends to be either beneficial (as with the fitness function used here) or detrimental, and bias the generation of new actions accordingly.

\subsection{The Artificial World}
These experiments used a default artificial world: a toroidal lattice with 1024 cells each occupied by a single, stationary agent, and a von Neumann neighborhood structure. Creators and imitators were randomly dispersed.\footnote{In other experiments (Leijnen \& Gabora, 2009a) we investigated the results of clustering creators.} Runs lasted 100 iterations, and all data are averages across 100 runs.

\subsection{A Typical Run}
Fitness and diversity of actions are initially low because all agents are initially immobile, implementing the same action, with all body parts in the neutral position. Soon some agent invents an action that has a higher fitness than immobility, and this action gets imitated, so fitness increases. Fitness increases further as other ideas get invented, assessed, implemented as actions, and spread through imitation. The diversity of actions increases as agents explore the space of possible actions, and then decreases as agents hone in on the fittest actions. Thus, over successive rounds of invention and imitation, the agents' actions improve. EVOC thereby models how adaptive change accumulates over time in a purely cultural context. 

\section{Experiment One: Effect of Varying the Ratio of Creators to Imitators}

The first experiment investigated how varying the level of creativity of individuals affects the fitness of ideas in society as a whole. To incorporate individual differences in degree of creativity we modified EVOC such that agents spanned the full range of possibilities from always creating, to always imitating, to in-between strategies in which agents created in some iterations and imitated in others. Those that could create at all are referred to as \emph{creators}. Those that only obtain new actions by imitating neighbors are referred to as \emph{imitators}. It was possible to vary the probability that creators create versus imitate (i.e., they range from creating all the time to behaving almost like imitators). Whereas any given agent is either a creator or an imitator throughout the entire run, the proportion of creators creating or imitating in a given iteration fluctuates stochastically.

\subsection{Procedure}
The proportion of creators relative to imitators in the society is referred to as $C$. The creativity of the creators---that is, the probability that a creator invents a new action instead of imitating a neighbor---is referred to as $p$. If a creator decides to create on a particular iteration, there is a 1/6 probability of changing the position of any body part involved in an action\footnote{This gave on average a probability of one change per newly created action, which previous experiments (Gabora, 1995) showed to be optimal.} The society consists of three subgroups:\\
\begin{APAenumerate}
\item {$C \times p \times N$} creators attempting to create\\
\item {$C \times (1-p) \times N$} creators attempting to imitate\\
\item {$(1-C) \times N$} imitators attempting to imitate\\
\end{APAenumerate}

In previous investigations we measured the diversity of ideas over the course of a run for different values of $C$ and $p$. We found that the cultural diversity, \emph{i.e.,} the number of different ideas implemented by one or more agent(s), was positively correlated both with the proportion of creators to imitators, and with how creative the creators were. We also obtained suggestive evidence that when creators are relatively uncreative, the mean fitness of ideas increases as a function of the percentage of creators in the society, but when creators are highly creative, the society appears to be better off with fewer creators (Leijnen \& Gabora, 2009b). However, this study had shortcomings. First, the simulations were performed with small societies of only 100 agents. Second, since action fitness was obtained at only one time slice (the 50th iteration) for all ratios of creators to imitators, these results did not reflect the dynamics of the time series. Given a set of series of accumulating value over time, it is unclear which series is most representative. The series cannot be unambiguously ordered unless for each pair of series one strictly dominates the other, and that is not the case here; the curves representing mean fitness at different values of $\{C, p\}$ increase monotonically but they may cross and re-cross as time progresses. Thus here we present a more extensive investigation of the relationship between creativity and society as a whole that employs a sophisticated solution to the time series problem.

\subsubsection{Analysis}
We used time series discounting which associates a ``present value'' with any future benefit such that the present value of any given benefit diminishes as a function of elapsed time until the benefit is 
realized (McDonald \& Siegel, 1986). The standard approach in financial settings is exponential discounting. Given a series of benefits $b_{t}$, the Net Present Value (NPV) is defined as:

\begin{equation}
NPV(b) =  \displaystyle\sum_{t=1}^N r^{t-1}  b_{t} \quad with \quad 0 < r \leq 1
\label{eq:npv}
\end{equation}

The discount rate $r$ is normally set as $r = (\frac{100+i}{100})^{-1}$ where $i$ is the interest rate (in percentage) for the unit period that an investor can obtain from a safe investment. 
This basic idea was adapted to analyze the benefit accrued by attaining fit actions for different values of $C$ and $p$ in EVOC. The first discounting method used was Time-to-Threshold (TTT) discounting. Since all fitness trajectories were monotonically increasing, those that reached a reasonably high threshold $\tau$ sooner should be valued higher. We measured how many iterations (time to threshold) it took for fitness to reach $\tau$. For these runs, $\tau = 9$ was used as a measure of optimal fitness to allow for a realistic averaging over time. 

Whereas imitators need creators, creators should ignore others if they could do better on their 
own ($p = 1$). In other words, the fitness prospects of creators' ideas when they work alone can be viewed in a manner analogous to the interest yield of treasury bonds in investment decisions. This logic suggests another kind of modification of the standard discounting method. The second adaptation to the basic notion of discounting we refer to as Present Innovation Value (PIV) discounting. Let $N$ be the number of iterations and let $F_{t}^{C,p}$ be the mean action fitness at iteration $t$ for parameter 
setting $\{C, p\}$. Thus $F_{t}^{1,1}$ is the fitness expectation with no interaction amongst agents. We define the PIV for any fitness curve as:
\begin{equation}
PIV(F^{C,p}) = - N + \displaystyle\sum_{t=1}^N \frac{F^{C,p}}{F^{1,1}}
\label{eq:piv}
\end{equation}

\noindent Thus the PIV value gives us a measure of the extent to which the mean fitness of outputs benefits or suffers as imitation becomes more prevalent (due to an increase in either the proportion of imitators or the probability that creators imitate) compared to a society composed solely of creators creating all the time with no imitation.

\subsection{Results and Discussion}
\begin{figure}
\centering
\includegraphics[width=\columnwidth]{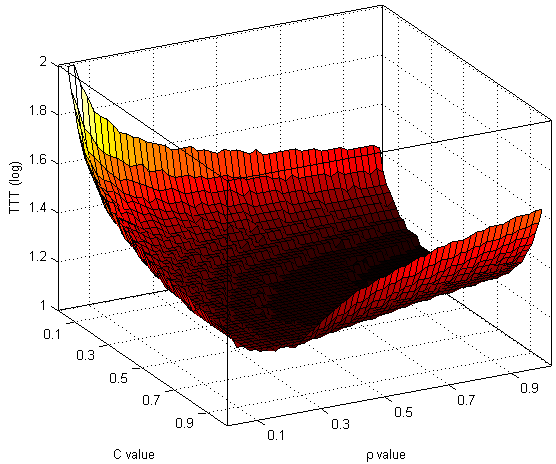}
\caption{3D graph of the log$_{10}$ Time-to-Threshold (TTT) landscape of the average mean fitness for different values of $C$ and $p$, with $\tau = 9$. The valley in the fitness landscape indicates that the optimal values of $C$ and $p$ for the society as a whole are less than their maximum values for most ${C,p}$ settings.}
\label{fig:TTT-cropped}
\end{figure}

\begin{figure}
\centering
\includegraphics[width=\columnwidth]{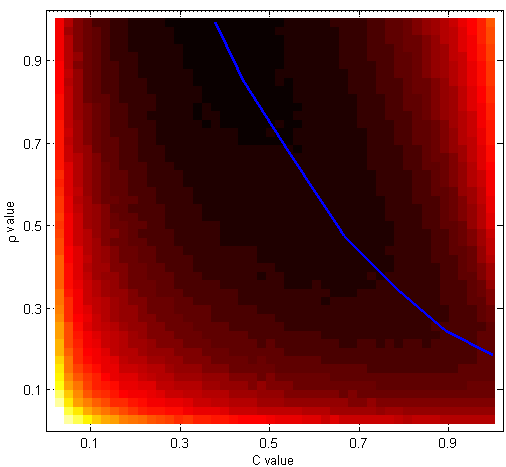}
\caption{Top-view contour plot of the log$_{10}$ Time-to-Threshold (TTT) landscape of the average mean fitness for different values of $C$ and $p$, with $\tau = 9$. The line, obtained by visually extrapolating over minimum values $C$ and $p$, indicates the set of optima.}
\label{fig:TTT-contour-line-cropped}
\end{figure}

All results are averages across 100 runs. The 3D graph and contour plot for the log$_{10}$ TTT discounting analysis of the time series for different ${C,p}$ settings are shown in Figures \ref{fig:TTT-cropped} and \ref{fig:TTT-contour-line-cropped}, respectively. Note that by definition a low TTT value corresponds to high mean fitness of actions across the society. The TTT method clearly demonstrates a valley in the adaptive landscape. The line running along the bottom of the valley in Figure \ref{fig:TTT-cropped} indicates, for any given value of $p$ the optimal value for $C$, and \emph{vice versa}. When $p = 1$ the optimal value of $C = 0.38$. When $C = 1$ the optimal value of $p$ is 0.19. The global optimum is at approximately $\{C,p\} = \{0.4,1.0\}$. 

\begin{center} 
Insert Figure \ref{fig:TTT-cropped} here.
\end{center}

\begin{center}
Insert Figure \ref{fig:TTT-contour-line-cropped} here.
\end{center}

\begin{figure}
\centering
\includegraphics[width=\columnwidth]{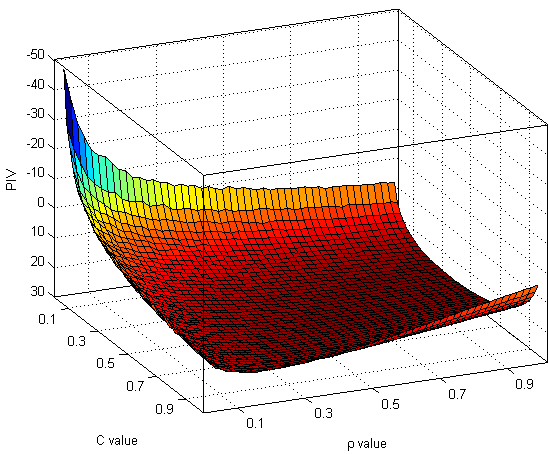}
\caption{3D graph of the Present Innovation Value (PIV) landscape of the average mean fitness for different values of $C$ and $p$. Since the $x$ axis has been inverted to aid visibility of the adaptive landscape, the valley again indicates that the optimal values of $C$ and $p$ for the society as a whole are less than their maximum values for most ${C,p}$ settings.}%
\label{fig:PIV-cropped}
\end{figure}

\begin{figure}
\centering
\includegraphics[width=\columnwidth]{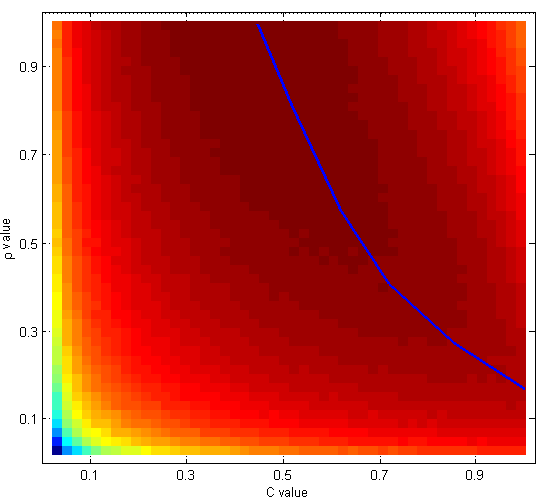}
\caption{Top-view contour plot of the Present Innovation Value (PIV) landscape of average mean fitness for different values of $C$ and $p$. The line, obtained by visually extrapolating over maximum values $C$ and $p$, indicates the set of optima.}
\label{fig:PIV-contour-line-cropped}
\end{figure}

The 3D graph and contour plot for the PIV discounting analysis of the time series for different ${C,p}$ settings are shown in Figure \ref{fig:PIV-cropped} and Figure \ref{fig:PIV-contour-line-cropped} respectively. The pattern is very  similar to that obtained with the  log$_{10}$ TTT discounting analysis. 

\begin{center}
Insert Figure \ref{fig:PIV-cropped} here.
\end{center}

\begin{center} 
Insert Figure \ref{fig:PIV-contour-line-cropped} here.
\end{center}

These results show that the first hypothesis---that society as a whole can suffer if either (1) the ratio of creators to imitators is too high, or (2) creators are too creative---was supported. Both log$_{10}$ TTT  and PIV analysis of the time series showed that, although some creativity is essential to get the fitness of cultural novelty increasing over time, more creativity is not necessarily better. For optimal mean fitness of agents' actions across the society there is a tradeoff between $C$, the proportion of creators in the artificial society, and $p$, how creative these creators are. 

\section{Experiment Two: The Effect of Social Regulation}
 The second experiment tested the hypothesis that society as a whole benefits when individuals can vary how creative they are in response to the perceived effectiveness of their ideas. In theory, if effective creators create more, and ineffective creators create less, the ideas held by society should collectively evolve faster. 

\subsection{Procedure}
Social regulation (SR) was implemented by increasing the invention-to-imitation ratio for agents that generated superior ideas, and decreasing it for agents that generated inferior ideas. To implement
this the computer code was modified as follows. Each iteration, for each agent, the fitness of its current action relative to the mean fitness of actions for all agents at the previous iteration was assessed. Thus we obtained the relative fitness, $RF$, of its cultural output. The agent's personal probability of creating, $p(C)$, was modified as a function of $RF$ as follows:

\begin{equation}
	p(C)_{n} = p(C)_{n-1} \times RF_{n-1}
\label{eq:socialregulation}
\end{equation}
 
The probability of imitating, $p(I)$, was 1 - $p(C)$. Thus when $SR$ was on, if the relative fitness of an agent's ideas was high the agent invented more, and if it was low the agent imitated more. $p(C)$ was initialized at 0.5 for both SR and non-SR societies. We compared runs with SR to runs without it. In this set of experiments only simple, single-step actions were possible. 

\subsection{Results and Discussion}
\begin{figure}
\centering
\includegraphics[width=0.95\columnwidth]{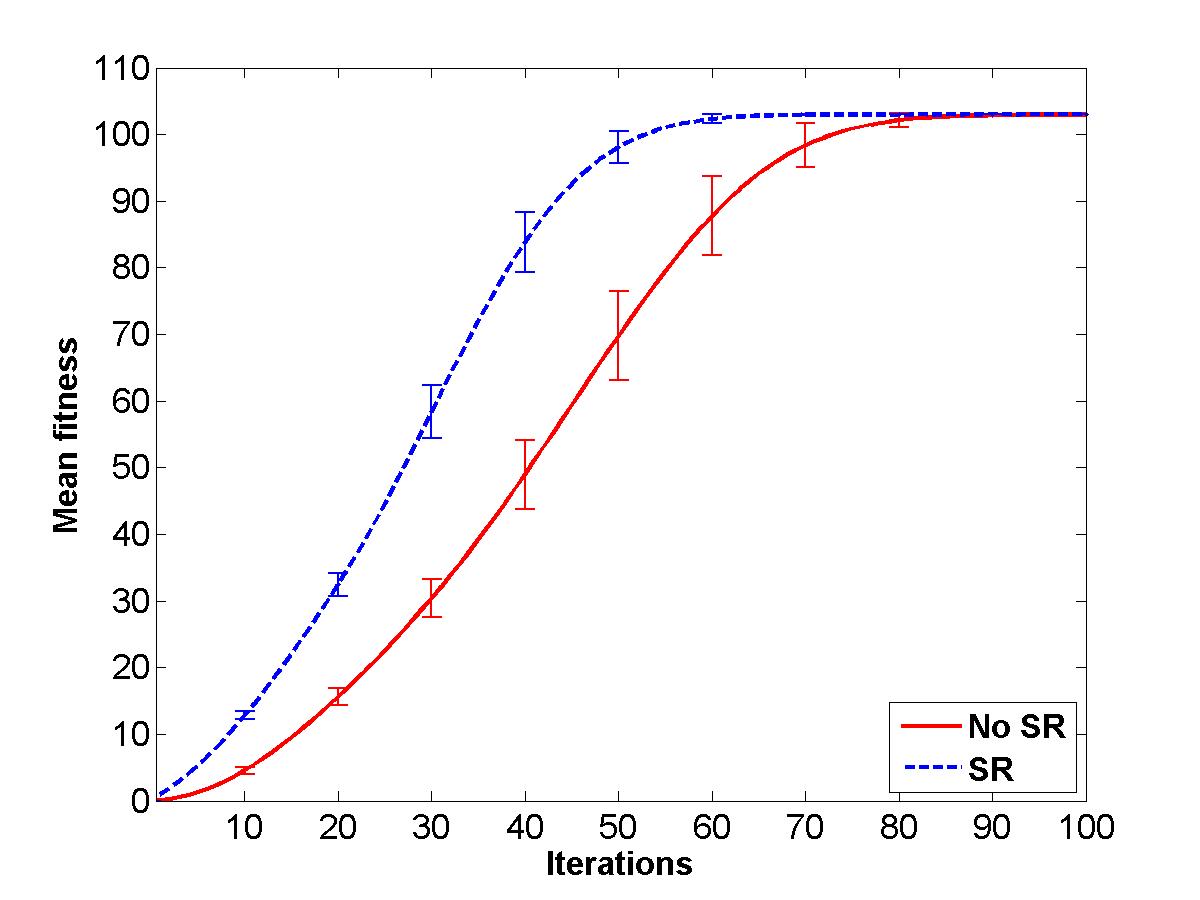}
\caption{This graph plots the mean fitness of implemented actions across all agents over the duration of the run, with and without social regulation.}
\label{fig:F1-no-CF-no-Chaining-fitness}
\end{figure}

The mean fitness of the cultural outputs of societies with SR (the ability to self-regulate inventiveness as a function of inventive success) was higher than that of societies without SR, as shown in Figure \ref{fig:F1-no-CF-no-Chaining-fitness}. Thus, these results show that the second hypothesis---that a society can perform better if individuals are able to adjust how creative they are over time in accordance with their perceived creative success---was also supported. However, the difference between SR and non-SR societies was temporary; the gap between them closed once the space of possible ideas had been explored. In both SR and non-SR societies mean fitness of actions plateaued when all agents converged on optimally fit ideas. Thus, the value of segregating into creators and imitators was short-lived.

\begin{center} 
Insert Figure \ref{fig:F1-no-CF-no-Chaining-fitness} here.
\end{center}

\begin{figure}
\centering
\includegraphics[width=0.95\columnwidth]{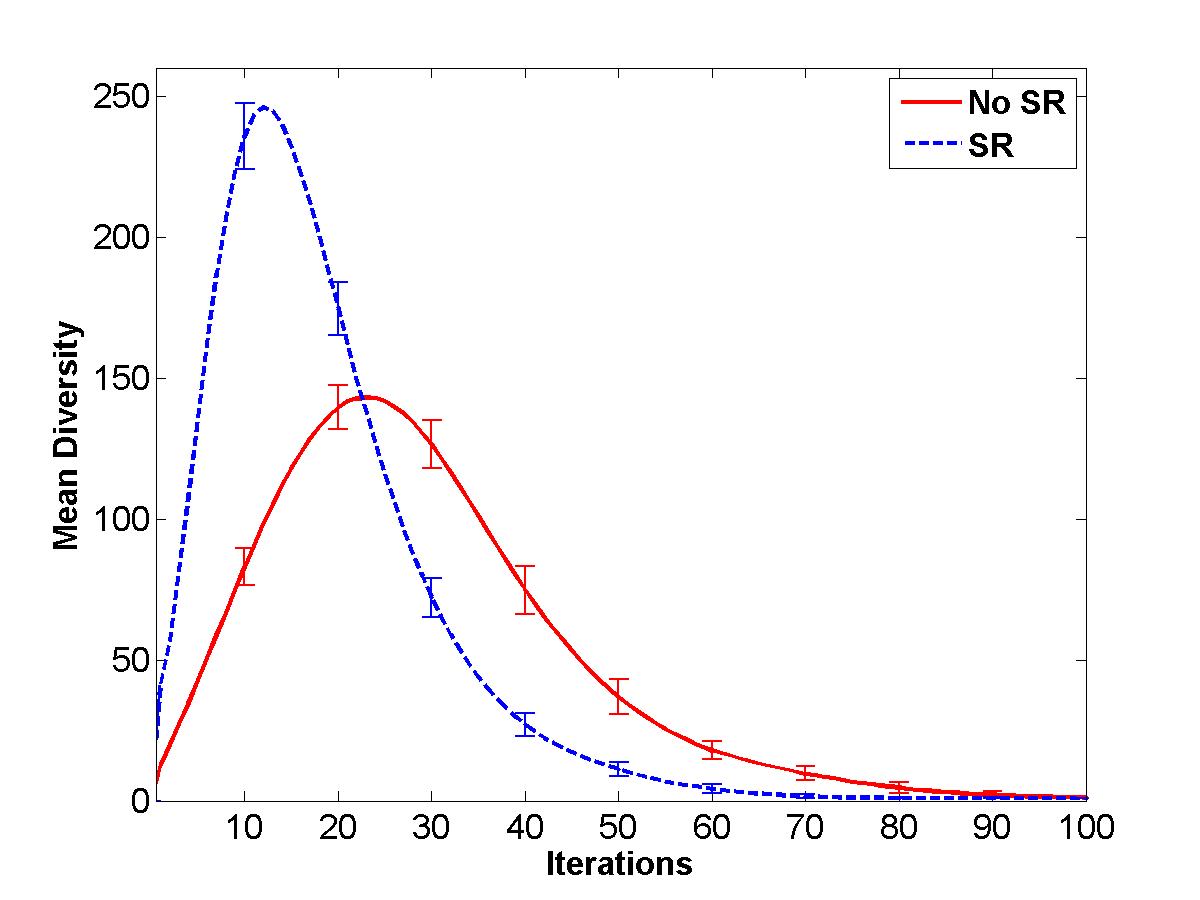}
\caption{This graph plots the mean diversity of implemented actions across all agents over the duration of the run, with and without social regulation.}
\label{fig:F2-no-CF-no-Chaining-diversity}
\end{figure}

The diversity, or number of different ideas, exhibited an increase as the space of possibilities was explored followed by a decrease as agents converged on fit actions, as shown in Figure \ref{fig:F2-no-CF-no-Chaining-diversity}. This diversity pattern is typical in evolutionary scenarios where outputs vary in fitness. What is of particular interest here is that this pattern occurred earlier, and was more pronounced, in societies with SR than in societies without it. With SR, superior creators were diverging in multiple directions, so making them more creative did increase diversity, while Inferior creators merely reinvent the wheel, so decreasing their creativity had little effect on the total number of different outputs. 

\begin{center} 
Insert Figure \ref{fig:F2-no-CF-no-Chaining-diversity} here.
\end{center}

\begin{figure} 
\centering
\includegraphics[width=0.52\columnwidth]{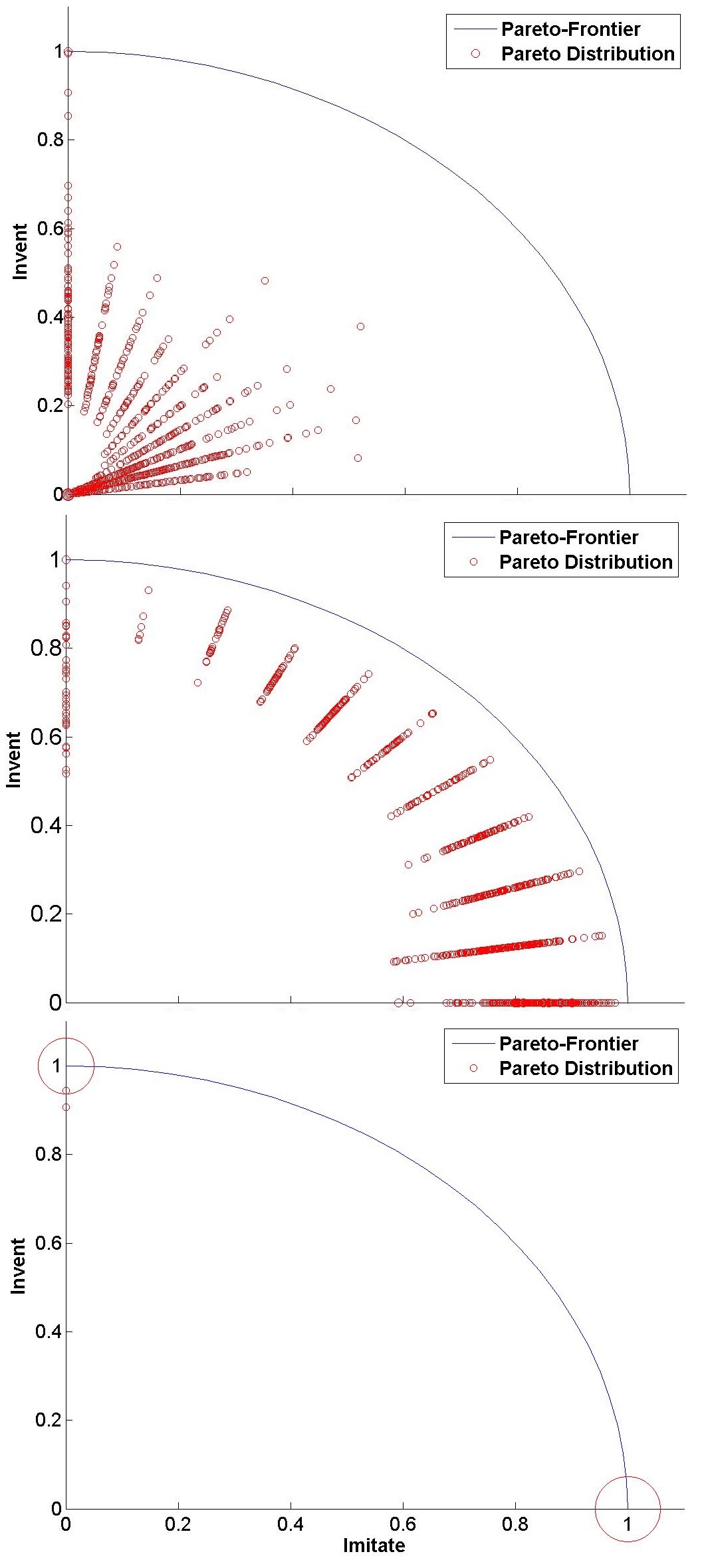}
\caption{This graph plots the fitness of actions obtained through invention on the y axis and through imitation on the x axis. Fitness values are given as a proportion of the fitness of an optimally fit action. The curved line is a pareto frontier because it consists of different optimal allocations of actions, ranging from always inventing optimally (upper left end of curve), to always implementing an optimal action obtained through imitation (bottom right end of curve), as well as strategies involving a mixture of inventing and imitating (all other points along the curve). Points to the left of this curve indicate strategies that involve the execution of suboptimal actions. Each small red circle shows the mean fitness of an agent's actions obtained through invention and imitation averaged across ten iterations: iterations 1 to 10 in the top graph, 25 to 35 in the middle graph, and 90 to 100 in the bottom graph. Since by iteration 90 all values were piled up in two spots---the upper left and the bottom right---they are indicated by large red circles at these locations.}
\label{fig:F3-agent-distribution}
\end{figure}

Although all agents initially invented and imitated with equal frequency, societies with SR ended up separating into two distinct groups: one that almost exclusively invented, and one that almost exclusively imitated, as illustrated in Figure \ref{fig:F3-agent-distribution}. Thus, the effect of SR on the fitness and diversity of outputs can indeed be attributed to increasingly pronounced individual differences in their degree of creativity over the course of a run. Agents that generated superior cultural outputs had more opportunity to do so, while agents that generated inferior cultural outputs became more likely to propagate proven effective ideas rather than reinvent the wheel.

\begin{center} 
Insert Figure \ref{fig:F3-agent-distribution} here.
\end{center}

\section{Experiment Three: Social Regulation with Chaining}

The short-lived but encouraging results of experiment two inspired experiment three, which tested the hypothesis that benefit from this social regulation mechanism could be longterm if the space of possible ideas were open-ended. The space of possible ideas was made open-ended by allowing simple ideas to be combined  or ``chained'' together into more complex ideas. Thus, over iterations the complexity of inventions could steadily increase. 

\subsection{Procedure}
The fitness function used in the previous experiments was only useful for single-step actions; once an agent found an optimal cultural output it continued to do the same thing, so eventually the mean fitness of actions across the society reached a plateau. In this next experiment, the chaining of simple actions into complex actions allowed for a potentially infinite variety of actions and no limit on their fitness.
 
To implement chaining it was necessary to modify the fitness function. We needed a fitness function that discouraged simply executing the same fit action again and again (to capture that cultural evolution entails the learning of sequences of different actions), that included a natural means of determining when a multi-part action would terminate, and that was conducive to the cultural evolution of actions that build cumulatively on previously learned or created actions. This was made possible using templates to constrain the space of allowable sub-actions that together constitute a complete action, using an adaptation of the Royal Roads fitness function (Forrest \& Mitchell, 1993). Definitions of terms used in the evaluation of the fitness of an action are provided in Table One. 

\begin{center}
Insert Table 1 here.
\end{center}

\begin{table}
\caption{Definitions used in the evaluation of chained actions.}
 \label{tab:definitiontable}
\setlength{\tabcolsep}{1em}
\setlength{\extrarowheight}{6pt}
\centering
 	\begin{tabular}{>{\raggedright}p{2.50cm}>{\raggedright}p{5cm}>{\raggedright\arraybackslash}p{5cm}}
	\toprule
	\hline
  Term & Definition & Example \\
  \midrule
   Body Part & \nohyphens{Component of agent other than neural network.} & Left Arm (LA) \\
   \nohyphens{Sub-action} & \nohyphens{Set of six components that indicates position of 6 body parts. Each can be in a neutral (0), up (1), or down (-1) position.} & \{HD:0, LA:1, RA:-1, LL:1, RL:0, HP:-1; \nohyphens{This sub-action is abbreviated 01-110-1\}}\\
   \nohyphens{Action} &  \nohyphens{One or more sequential sub-actions.} & \{\{01001-1\}, \{-10-1-111\}\}  \\
 \nohyphens{Template} & \nohyphens{Abstract or prototypical format for a sub-action. Position of a body part can be unspecified (*).} & \{HD:0, LA:*, RA:1, LL:*, RL:1, HP:-1\} \\
\hline
 \bottomrule
 \end{tabular}
\end{table}

The fitness function was determined by 45 templates. The templates can be thought of as defining the cultural significance or utility of types of sub-actions (such as dance steps). Each template $T^i$ consists of six components, one for each body part (\emph{i.e.,} $T^i={t_j^i };j=1..6$). Each body part can be in a neutral position (0) , up (1), down (-1), or an unspecified position (*). Six examples of templates are provided in Table Two. For example, in template $T^i={*,1,-1,*,*,0}$, the left arm is up (LA:1), the right arm is down (RA:-1), the hips are in the neutral position (HP:0), and the positions of other body parts is unspecified (HD:*, LL:*, and RL:*). The templates provide constraints, as well as flexibility with respect to what constitutes a fit action. For example, in an optimally fit action, the head must be in the neutral position (in $T^1$ the first component is 0) but the positions of other body parts can vary). 

\begin{center}
Insert Table 2 here.
\end{center}

\begin{table}
\caption{A partial set of the templates used in the first fitness function}
 \label{tab:templatestable}
\setlength{\tabcolsep}{1.5em}
\setlength{\extrarowheight}{6pt}
\centering
  \begin{tabular}{ll}         
	\toprule 
	\hline
		$T^1=\{0,*,*,*,*,*\}$ & $T^{24}=\{1,*,*,1,1,*\}$  \\ 
		$T^2=\{*,0,*,*,*,*\}$ & $T^{25}=\{1,*,1,*,1,*\}$   \\ 
		$T^3=\{*,*,0,*,*,*\}$ & $T^{26}=\{1,*,1,1,*,*\}$  \\ 
		
\hline 		
\bottomrule
\end{tabular}
\end{table}

\subsubsection{Calculating the fitness of a template}
Assume that D is a sub-action (\emph{i.e.,} $D={d_j};j=1..6$) and $T^i$ is the $i^{th}$ template (\emph{i.e.,} $T^i={t_j^i};j=1..6$). Thus, $d_j$ represents the position of the $j^{th}$ body part and the value of $d_j$ can be either 0 (neutral), 1 (up), or -1 (down). Likewise, the value of $t_j^i$ can be 0, 1, -1, or * (unspecified). Accordingly, the fitness of sub-action D is obtained as follows:

\begin{equation}
	F(D)=\sum_{i=1}^{19}{\Phi(T^i,D) \times \Omega(T^i)}
\label{eq:F D}
\end{equation}

As shown in this equation, fitness is a function of template weight ($\Phi(T^i,D)$) and template order ($\Omega(T^i)$).

$\Phi(T^i,D)$ is a function that determines the weight of sub-action $D$ by comparing it with template $T^i$. This weight is set to one if each component of the sub-action (\emph{i.e.,} $d_j;j=1..6$) either matches the corresponding component of the template (\emph{i.e.,} $t_j^i;j=1..6$) or if the corresponding components of the template is unspecified (\emph{i.e.,} $t_j^i= *$), thus:
\begin{eqnarray}
	\Phi(T^i,D) = \bigg\{ 
	\begin{matrix}
		1 & if \ \forall t_j^i \in T^i:t_j^i= d_j \ or \ * \\ 0 & otherwise	
	\end{matrix}	
\label{eq:template weight}
\end{eqnarray}

$\Omega(T^i)$ computes the order of the template $T^i$ by counting the number of components that have a specified value (\emph{i.e.,} $t_j^i \neq *)$.
\begin{equation}
\Omega(T^i)= \sum_{j=1, t_j^i \neq *}^6{t_j^i}
\label{eq:template order}
\end{equation}

\bigskip \noindent The 
acceptable sub-actions are $\{0,1,-1,1,-1,1\}$, $\{0,1,-1,1,-1,-1\}$, $\{0,-1,1,-1,1,1\}$, and  $\{0,-1,1,-1,1,-1\}$.

\bigskip \indent The fitness function is difficult to solve because it is rugged; there are multiple milestones, or fitness peaks, that agents must achieve before reaching a plateau. For example, in Table 2 we see that the action {0,0,0,0,0,0} has a fitness of 6. An agent may move on from this action to find an action that fits the third order templates with a fitness of 31, {\it e.g.,} $F(D):\{1,1,1,1,1,0\}=3+3+3+3+3+3+3+3+3+3+1=31$. 

\bigskip 
\subsubsection{Modeling chaining}
The chaining algorithm is illustrated schematically in Figure 9.
Chaining gives agents the opportunity to execute multi-step actions, thereby increasing the potential diversity of actions and making the space of possible actions open-ended. An agent can keep adding a new sub-action to its current action so long as the most recently-added sub-action is both novel and successful. 

\bigskip 
\begin{center}
Insert Figure 9 here. 
\end{center}

\begin{figure} 
\linespread{2}
\centering
\includegraphics[width=0.95\columnwidth]{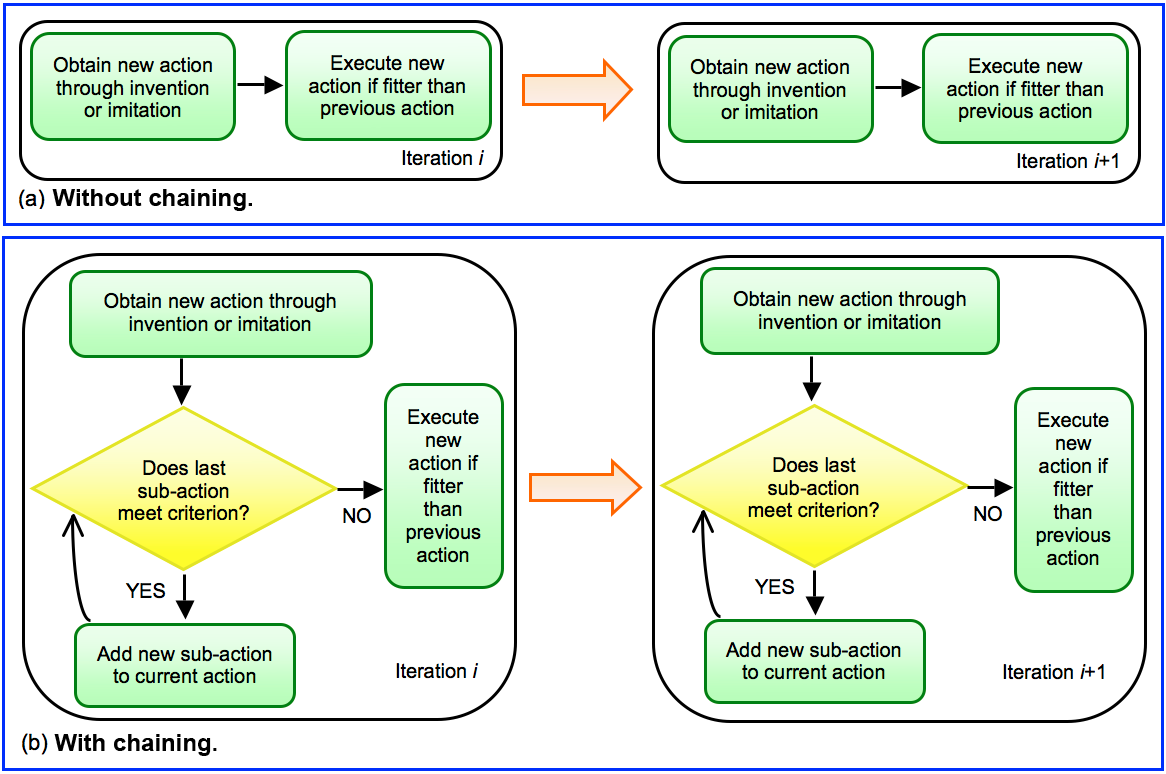} \\
 \begin{center} 
\caption{Schematic illustration of the process by which an agent determines what action it will implement in the next iteration without chaining (above) and with chaining (below).}
 \end{center} 
\label{fig:chaining-fig}
\end{figure}

A sub-action D is considered novel if at least one of its components is different from that of the previous sub-action. This ensures that a multi-part action actually consisted of multiple parts (rather than a drawn-out execution of the same sub-action). It is considered successful if there exists a template $T^i$ such that $\Phi(T^i,D)$ is one:

\begin{equation}
successful(D) = \bigg\{ \begin{matrix}
	true & if \ \exists \ T^i : \Phi(T^i,D) = 1 \\
	false & otherwise 
\end{matrix}
\label{eq:chaining}
\end{equation}


\bigskip \noindent The ``successful'' constraint was added to mimic the fact that real human actions such as gesturing and tool-making are generally highly constrained. 



The fitness $F_c$ of a multi-step action with $n$ chained single-step actions (each with fitness $F_n$) is calculated as follows:

\begin{equation}
F_c= \sum\limits_{k = 1}^n {F_n} 
\label{eq:fitnesschained}
\end{equation}

\bigskip \noindent Thus agents could execute multistep actions, and the optimal way of going about any particular step depended on how one went about the previous step. So long as the agent continued to invent acceptable new sub-actions, an action could be arbitrarily long. In general, the more sub-actions the fitter the action. This is admittedly a simple way of simulating the capacity for chaining, but we were not interested in the impact of these actions {\it per se}. The goal here was simply to enable create a world in which improvement is always possible. 

Note that since multi-step actions tended to be fitter than single-step actions there was a bias towards multi-step actions. This was necessary to test the hypothesis that SR is only advantageous so long as it is possible to obtain fitter outputs than those currently in use. This aspect of the model seems fairly realistic; new ideas do tend to build on old ones, and often involve increasingly more steps to achieve their final form, and these new multi-step ideas are often (though not always) fitter than what came before. 

\subsection{Results and Discussion}

\begin{figure}
\centering
\includegraphics[width=0.95\columnwidth]{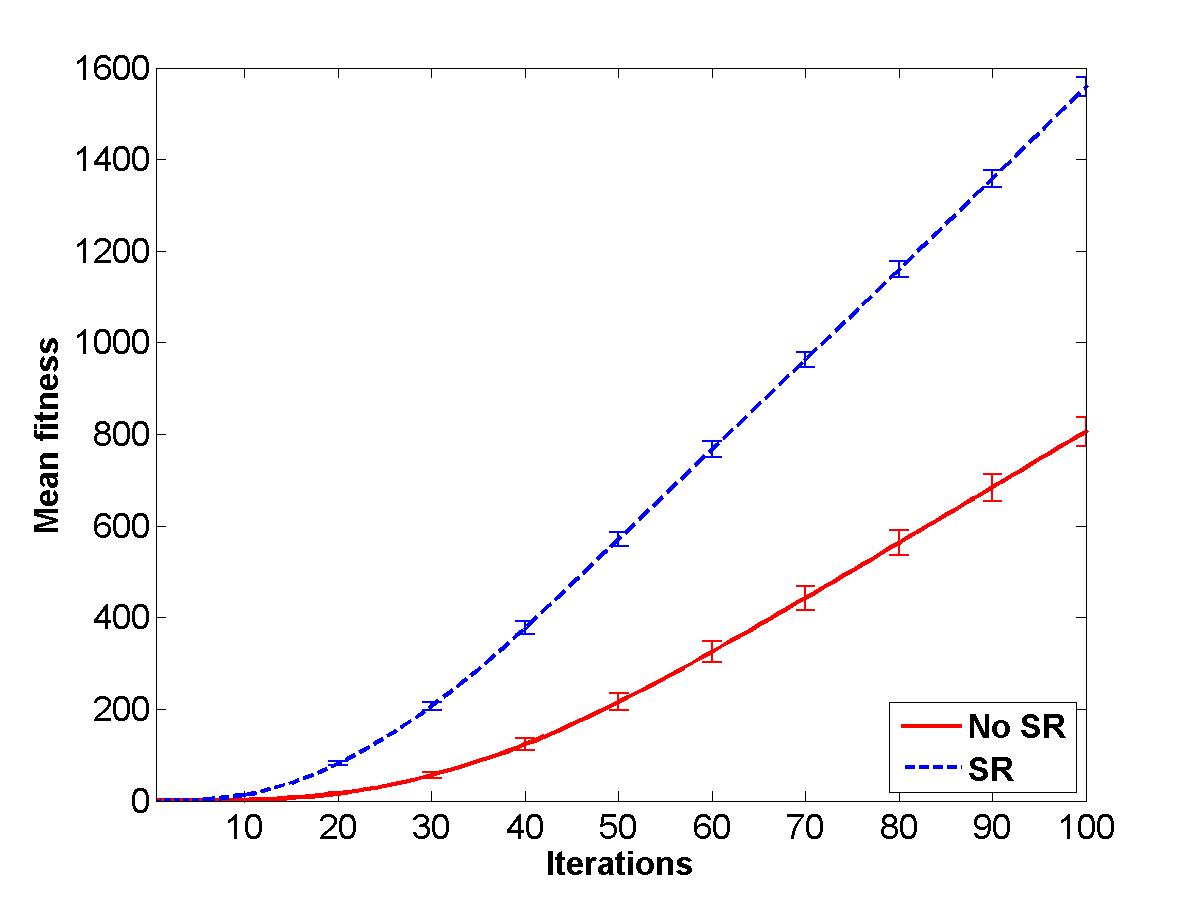}
\caption{This graph plots the mean fitness of actions across all agents over the course of the run with chaining turned on, with and without social regulation (SR).}
\label{fig:F4-no-CF-yes-Chaining-fitness}
\end{figure}

With chaining turned on, cultural outputs became increasingly fit over the course of a run, as shown in Figure \ref{fig:F4-no-CF-yes-Chaining-fitness}. This is because a fit action could always be made fitter by adding another sub-action. 
Thus the third hypothesis---that in order for the benefit of this social regulation mechanism to be ongoing (as opposed to temporary), the space of possible creative outputs must be open-ended, such that it is always possible for superior possibilities to be found---was supported.

\begin{center} 
Insert Figure \ref{fig:F4-no-CF-yes-Chaining-fitness} here.
\end{center}

\begin{figure}
\centering
\includegraphics[width=0.95\columnwidth]{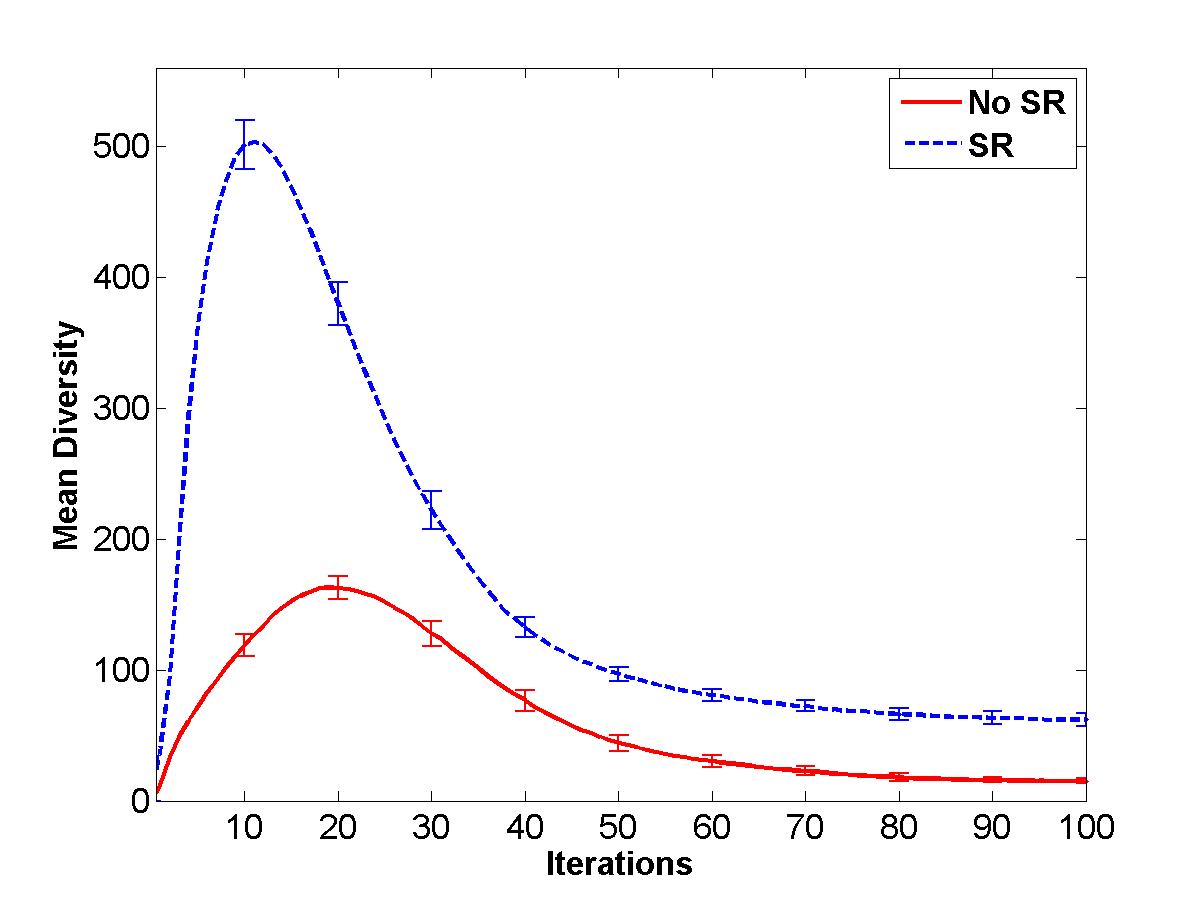}
\caption{This graph plots the  mean diversity of implemented actions across all agents over the course of the run with chaining, with and without social regulation (SR).}
\label{fig:F5-no-CF-yes-Chaining-diversity}
\end{figure}

As was the case without chaining, the diversity of ideas with chaining exhibited an increase as the space of possibilities was explored, followed by a decrease as agents converged on fit actions, and once again the peak in diversity is earlier and more pronounced with SR than without it, as shown in Figure \ref{fig:F5-no-CF-yes-Chaining-diversity}. 
However SR diversity remains higher than non-SR diversity throughout the run because the agents did not converge on a static set of actions; their actions changed continuously as they found new, fitter actions. Moreover, SR runs contain creators that are executing highly complex actions, and there are more ways of executing a complex action than a simple one.

\begin{center} 
Insert Figure \ref{fig:F5-no-CF-yes-Chaining-diversity} here.
\end{center}

\begin{figure}
 \vspace{1\baselineskip} 
\centering 
\includegraphics[width=0.50\columnwidth]{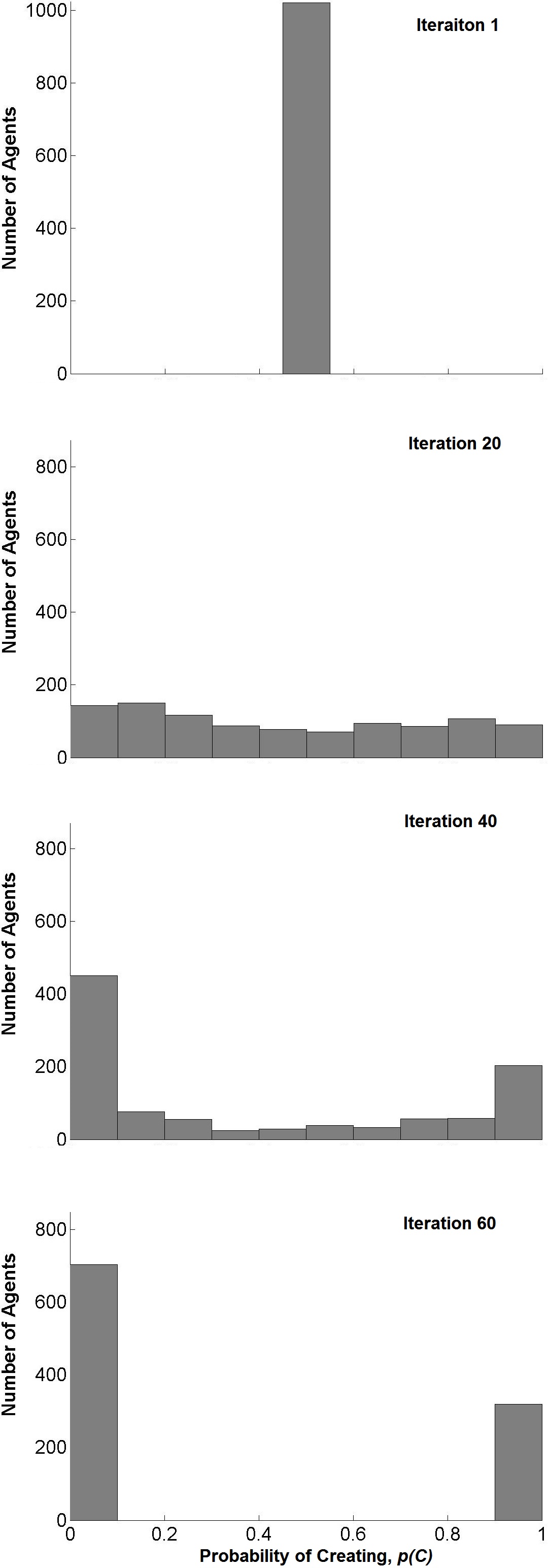}
\caption{At the beginning of the run (top) all agents created and imitated with equal probability. Midway through the run their $p(C)$ values were distributed along the range of values from 0 to 1. By the end of the run (bottom) they had segregated into imitators (with $p(C)$ from 0 to 0.1) and creators (with $p(C)$ from 0.9 to 1).}
\label{fig:0figdistribution}
\end{figure}

Once again we know that the effects of SR on mean fitness and diversity were due to the segregation of agents over time into distinct groups: those who almost exclusively invented and those who almost exclusively imitated, as illustrated in Figure \ref{fig:0figdistribution}. Since with SR there were increasingly pronounced individual differences in degree of creativity over the course of a run, the differences between SR and non-SR societies can indeed be attributed to the fact that the best creators were not wasting iterations trying to imitate inferior neighbors, they could reach relatively remote and complex ideas more quickly. Agents that generated superior cultural outputs had more opportunity to do so, while agents that generated inferior cultural outputs became more likely to propagate proven effective ideas.

\begin{center}
Insert Figure \ref{fig:0figdistribution} here.
\end{center}

Figure \ref{fig:0figdistribution} shows that when some agents start to specialize in creating, others start to specialize in imitating, such that across the society as a whole the balance between creating and imitating is maintained. Bear in mind that since in all the simulations reported here the agents are stationary and can only imitate immediate neighbors, it is not the case that imitators are just imitating themselves. In other words, creators and imitators are not segregated spatially, and there is transmission between them. Thus the imitators' efforts are, indirectly, playing a role in the generation of novelty. 

Thus, as the balance between creating and imitating gets tilted one way or the other within individual agents, a new kind of between agents balancing act starts to unfold, such that both the generation and proliferation of novelty are preserved. 
These results support the hypothesis that it is algorithmically possible for social regulation of individual creativity levels to be a means by which a society balances novelty with continuity. To our knowledge, this hypothesis has never before been put forward, let alone tested. The results suggest that it would be fruitful to investigate whether in real human societies local exchange of social cues regarding the desirability of creative efforts has the global effect of balancing novelty with continuity. 

\section{General Discussion}
%
%
The experiments reported here were inspired in part by early work in evolutionary theory showing that evolution entails a synthesis of processes such as mutation that \emph{generate variants}, and processes such as heredity that \emph{preserve fit variants} (Haldane, 1932). Our results suggest that the generation of cultural novelty through creative processes is tempered by social learning processes that preserve fit ideas, and that achieving a delicate balance between the two has benefits for society at large. Although EVOC agents are highly rudimentary, the model incorporates a drawback of creating: it incurs costs in terms of time and foregone alternatives. When creative agents invest in new ideas at the expense of imitating proven ideas they effectively rupture the fabric of the artificial society by impeding the diffusion of tried-and-true solutions. Imitators, in contrast, serve as a ``cultural memory'' that ensure that valuable ideas are preserved. 
 
Experiment One tested the hypothesis that society as a whole can suffer if either (1) the proportion of creative individuals is too high, or (2) creative individuals are too creative, by carrying out a set of runs in an ABM that systematically varied the ratio of creators to imitators, and how creative the creators were. We observed a trade-off between these two variables, i.e., if there were few creators they could afford to be more creative, and vice versa, if there were many, their creativity had to be restrained to exert the same global benefit for society as a whole. 

Experiment One has intriguing though speculative implications for hiring practices in which individuals are expected to work in groups. It suggests that it may be productive to consider prospective employees in the context of existing team members and specifically where they stand on the creativity-conformity spectrum. If there are many creatives, or if they are extremely creative, it may be beneficial to balance the team with imitators, and vice-versa, to achieve the balance needed to hasten the cultural evolution of fit outputs. 

The results of Experiment One are consistent with a recent study of 155 ceramic tile companies in Spain, which found preliminary evidence of a saturation point beyond which excessive creativity interferes with proliferation of valuable designs, and thus decreases productivity (Vallet-Bellmunt \& Molina-Morales, 2015). This study also found that companies tended to do better if they focused on one of two alternative strategies: focus on creativity, or focus on centrality, the later of which entails accessing knowledge from other, related companies. They write, ``as both are resource and time-costly strategies, not only is a combination of them not synergic, but may in fact become negative for firms’ innovation performance.'' (p. 14) This provides preliminary evidence that a segregation into creator and imitator strategies need not necessarily occur at the level of individuals; it may occur at the level of companies, with adaptive consequences in the real world. 

Experiments Two and Three investigated another way in which creativity may be tempered with conformity: over time, those whose creative efforts are successful might increase their creativity while  those whose creative efforts are unsuccessful might decrease their creativity, and rely instead on social learning. 
Experiment Two tested the hypothesis  that a society as a whole can perform better if individuals are able to adjust how creative they are over time in accordance with their perceived creative success. When agents that were successful creators created more, and those that were unsuccessful creators created less (SR), the mean fitness of outputs was higher and the increase in diversity was more pronounced. Moreover, these results were due  to the segregation of agents over time into two groups---creators and imitators. Thus we showed that it is possible to increase the mean fitness of ideas in a society by enabling them to regulate how creative they are.

In Experiment Two the effect of SR on the pace of cultural evolution was short-lived. Since the space of possibilities was closed, all agents eventually converged on optimal outputs, and they could not find even better actions through chaining; there was a ceiling effect. At this point there was no longer any social benefit to having some members of society be dedicated creators. We hypothesized that this was because the agents were limited to a finite set of simple ideas. 

In Experiment Three, where the space of possible outputs was open-ended, and agents could execute multi-step outputs, agents once again segregated into creators and imitators. However, comparing Figure \ref{fig:F1-no-CF-no-Chaining-fitness} with Figure \ref{fig:F4-no-CF-yes-Chaining-fitness}, we see that there is no longer a ceiling effect; the difference in fitness between societies with SR and societies without it now increases throughout the run. 
Thus, the results support our third hypothesis that it is possible for the benefit of this social regulation mechanism to be ongoing rather than temporary, if the space of possible creative outputs is open-ended, such that it remains possible for fitter possibilities to be found. 
These findings suggest that it can be beneficial for a social group if individuals follow different developmental trajectories in accordance with their demonstrated successes, but only if the space of possible ideas is such that there are always avenues to explore for new creative ideas.

Although the simplicity of the model must be kept in mind before jumping to conclusions, the results of Experiments Two and Three are consistent with empirical findings concerning the value of imitation in collaborative groups, e.g., when people have access to their peers' solutions, imitation facilitates not just scrounging but the propagation of good solutions for further cumulative exploration  (Wisdom, Song, \& Goldstone, 2013). These results fit in well with evidence compiled by Florida (2002) that a natural distinction emerges in societies between the conventional workforce and the creative class. Our results further suggest that this division of labor is adaptive; when social regulation was in place the society as a whole benefited. The society was able to capitalize on both the creative abilities of the best creators and the dissemination of fit cultural outputs by the rest. This is in line with current thinking on the adaptive value of individual differences in personality across group members (Nettle, 2006).

These results do not prove that in real societies successful creators invent more and unsuccessful creators invent less; they merely show this kind of regulation of creativity at the individual level is a feasible means of increasing the mean fitness of creative outputs of the group as a whole. However, the fact that strong individual differences in creativity exist (Kaufman, 2003; Wolfradt \& Pretz, 2001) suggests that this does indeed occur in real societies. The question of whether, and how, social groups balance creativity with conformity would appear to be a promising area for future research. Based on the results obtained here, it seems reasonable to hypothesize that---whether the regulation is prompted by individuals themselves or mediated by way of social cues---families, organizations, or societies spontaneously self-organize to achieve a balance between creative processes that generate innovations and the imitative processes that disseminate these innovations. In other words, they evolve faster by tempering novelty with continuity. 

A more complex version of this scheme is that individuals find a task at which they excel, such that for each task domain there exists some individual in the social group who comes to be best equipped to explore that space of possibilities. To explore this in EVOC would have required an individualized or dynamically changing fitness function. Elsewhere we have investigated the effect of individualized and dynamically changing fitness functions on the fitness and diversity of cultural outputs (Gabora, 2008a; Gabora, Chia, \& Firouzi, 2013). Although for the present initial explorations the results were most easily interpretable with a static fitness function, in future research it would be interesting to investigate how individualized or dynamic fitness functions affect SR.

In real life, people are faced with numerous different tasks on a daily basis that are evaluated according to different criteria. Actions that are fit or appropriate for one purpose (e.g., for making a specific tool) are not necessarily fit for another purpose (e.g., for expressing agreement). Accordingly, if one examines only one creative task at a time, a society may appear to perform optimally when the creating is left to a subset of individuals. However, when one examines multiple creative tasks, the situation may be more complex; in the extreme, everyone would find a different specialized niche for their creative output, and be an imitator with respect to other specialized niches. Experiments with a model that is related to but very different from EVOC suggest that the capacity for hybrid learning---wherein agents acquire knowledge pertaining to one  environmental dimension through individual learning and knowledge pertaining to another environmental dimension through imitation---can foster specialization that benefits society as a whole (Kharratzadeh, Montrey, Metz, \& Shultz, 2015). Since creative problem solving is a form of individual learning, it seems reasonable to suggest there may be social benefits when individuals limit creative exploration to one or a few domains and for other domains rely on social learning. The value of this arrangement hangs on the extent to which creativity is domain-specific. The evidence here is mixed; although the capacity for expert-level creative achievement may be predominantly limited to a single domain (Baer, 1996; Ruscio, Whitney, \& Amabile, 1998; Tardif \& Sternberg, 1988), many if not most individuals may be able to experience personally meaningful and fulfilling creative engagement and express their personal creative style through multiple domains (Gabora, O'Connor, \& Ranjan, 2012; Hocevar, 1976; Plucker, 1998; Ranjan, 2014; Runco, 1987). The optimal distribution of creators and imitators across different tasks may be therefore be complex, with different individuals tending to specialize for different tasks, but some individuals exhibiting a generalized tendency toward creativity and others exhibiting a generalized tendency toward imitation. 

It has been suggested that the capacity to merge thoughts and ideas into chains of association or `streams of thought' initially emerged approximately 1.7 million years ago due to increased cranial capacity accompanying the transition from Homo habilis to Homo erectus (Donald, 1991). The increase in cranial capacity could have allowed for more fleshed out representations, which in turn allowed for more associative pathways amongst representations, and greater potential for streams of abstract thought. Mathematical (Gabora \& Aerts, 2009; Gabora \& Kitto, 2013) and computational (Gabora \& DiPaola, 2012; Gabora, Chia, \& Firouzi, 2013) models support the feasibility of this scenario. The fact that in the experiments reported here social regulation of creativity was found to be of lasting value only in societies composed of agents capable of chaining suggests that there may have been insufficient selective pressure for social regulation of creativity prior to onset of this capacity. Thus, individual differences in creativity would be expected to have emerged after this time.

The social practice of discouraging creativity until the individual has proven him- or herself may serve to ensure that creative efforts are not squandered. 
Individual differences in responsiveness to social cues may ensure that some percentage of society consists of individuals whose affiliative needs are low, and who therefore feel relatively free to deviate from social norms and be creative. Those individuals who are most tuned to social norms and expectations may over time become increasingly concerned with imitating and cooperating with others in a manner that promotes cultural continuity. Thus, their thoughts travel more well-worn routes, and they become increasingly less likely to innovate. Others might be tuned to the demands of creative tasks,  less tethered to social norms and expectations, and therefore more likely to see things from unconventional perspectives. Thus, they become more likely to come up with solutions to problems or unexpected challenges, find new avenues for self-expression, and contribute to the generation of cultural novelty. What Cropley et al. (2010) refer to as the ``dark side of creativity'' may to some degree reflect that the creative individual is tuned to task needs at the expense of human needs; ideas, not people, are the objects of their nurturing. Although in the long run this benefits the group as a whole because it results in creative outputs, in the short run the creative individual may be less likely to obey social norms and live up to social expectations, and to experience stigmatization or discrimination as a result, particularly in his/her early years (Craft, 2005; Scott, 1999; Torrance, 1963). Once the merits of such individuals' creative efforts become known, they may be supported or even idolized.

A limitation of this work is that EVOC in its current implementation does not accommodate selective or partial imitation. In other words, EVOC does not allow an agent to imitate some features of an idea and not others. An agent either copies exactly what a neighbor is doing or ignores that neighbor entirely for that iteration; it cannot choose bits and pieces that would augment or complement its own current action. Nor can an agent selectively combine elements of multiple different neighbors' actions at once. Consequently, imitation, while essential to the rapid spread of superior outputs, exacerbates convergence on a small set of solutions, i.e., it has a destructive effect on diversity. We expect that this effect would be reduced in investigations that incorporate partial imitation. Partial imitation would also be useful for dealing with what in biology is referred to as epistasis, wherein what is optimal with respect to one component depends on what is going on with respect to another. Once both components have been optimized in a mutually beneficial way (in EVOC, for example, symmetrical movement of both arms), excess creativity risks breaking up co-adapted partial solutions. Note that the goal of this paper was not to develop a realistic model of creativity per se but to investigate social factors in creativity. Nonetheless, in future studies we plan to increase the sophistication of the mechanisms by which agents create by incorporating ideas from the psychology of creativity (e.g., Gabora, 2017; Ward, Smith, \& Vaid, 1997) and formal models of individual creativity (e.g., Costello \& Keane, 2000; Dantzig, Raffone, \& Hommel, 2011; Thagard \& Stewart, 2011). 

There are other avenues for future investigation suggested by this work. One is to study more thoroughly the extent to which these findings are affected by the nature of the task, or neighborhood network structure (cf. Jacobs, 2000; Liu, Madhavan, \& Sudharshan, 2005). Another avenue for future research is to investigate the impact of varying the extent to which the generation of novelty is goal-directed versus random. An early experiment on a predecessor to EVOC (Gabora, 1995) investigated (1) the effect of turning on or off the ability to learn trends that bias the generation of subsequent novelty, and (2) the effect of varying the extent to which new ideas deviate from previous ideas. Both the ability to learn trends and the tendency to use successful known ideas as a basis for generating new ideas decrease the extent to which generation is random. The ability to learn trends increased the speed of convergence and decreased the diversity of ideas. Performance was optimal when, on average, new ideas deviated from old ones with respect to one component (i.e., movement of one body part changed). Either increasing or decreasing the extent to which new ideas deviated from old ones affected the speed of convergence but not the overall pattern of results. 

Yet another avenue for future research is to investigate the relative impact of nature versus nurture. In Experiment Three, the observed individual differences were completely due to `nurture' rather than `nature'. It would be interesting to see how initializing a run with individual differences in creative ability amongst agents affects the pattern of results. It would also be interesting to investigate the impact of individual differences in agents' responses to assessments of their creative efforts, such as variation in their responses to positively versus negatively valence assessments. This would enable us to, for example, model the  impact of possible gender differences in the tendency to decrease creative output in response to negative assessments of one's creative work, or increase creative output in response to positive assessments. If this is the case then even if the creative potential of boys and girls is initially equal they may exhibit gender differences in creativity by the time they reach adulthood. 

As mentioned in the introduction, many supposed models of cultural evolution are actually models of cultural \emph{transmission}. To demonstrate cultural transmission, as few as two cultural variants with error-prone copying is sufficient, whereas cultural evolution entails cumulative, creative, open-ended, adaptive cultural change. This confusion has led to misleading claims and analyses. Some of these are discussed elsewhere (Gabora, 2011, 2013); one that has not been discussed concerns Rogers’ (1988) paradox: the finding that when social learning and individual learning strategies are at equilibrium, social learning does not enhance average individual fitness. Although much as been made of Rogers' result (see Enquist, Eriksson, \& Ghirlanda, 2007; Kameda \& Nakanishi, 2003; Kharratzadeh, Montrey, Metz, \& Shultz, 2015; Rendell, Fogarty, \& Laland, 2010), his conclusions hinge on the assumption of a temporally varying environment, for without this there is no benefit in the model to individual learning. The value of individual learning here lies solely in that it facilitates the tracking of environmental change because \emph{the model does not incorporate creativity}. 

The supposed paradox yielded by Rogers' model reflects an underlying lack of understanding of the central role of creativity in individual learning. In the real world, even if the environment remains basically unchanged, we benefit from finding creative new ways of conceptualizing and responding to this world. Moreover, the distinction between social learning and individual learning may not be as fundamental as Rogers' model assumes to be; for example, it isn't obvious that imitating a peer is fundamentally different from imitating a cartoon character, or from a beatboxer imitating the sounds of instruments, or a dancer imitating the wind. The approach taken here speaks to the new and important questions and perspectives that can be addressed when creativity is incorporated into a model of cultural evolution.

\section{Conclusions}
While society seems to value creativity in the abstract, social institutions are often perceived as wielding excessive social pressure to conform, and placing obstacles to creative self-expression in the paths of creative individuals until they have proven themselves. The results of the experiments reported here suggest that there is a logic to these seemingly contradictory messages. Since a proportion of individuals benefit from creativity without being creative themselves by imitating creators, the rate of cultural evolution increases when the novelty-generating effects of creativity are tempered with the novelty-preserving effects of imitation. 
If there were few creators they could afford to be more creative, and vice versa; if there were many their creativity had to be restrained to exert the same global benefit for the society. Excess creativity was detrimental because creators invested in unproven ideas at the expense of propagating proven ones. 

We also obtained evidence that society can benefit by rewarding and punishing creativity on the basis of creative success. When each agent regulated its invention-to-imitation ratio as a function of the fitness of its cultural outputs, they segregated into creators and imitators, and the mean fitness of cultural outputs was higher. When the space of possible outputs was fixed, the beneficial effect of social regulation was temporary. However, making the space of possible outputs open-ended by enabling agents to chain simple outputs into complex ones, caused the social regulation induced increase in mean fitness of cultural outputs to be sustained. 

Although the model used here is vastly simpler than real societies it enabled us to manipulate the ratio of creators to imitators and the degree to which creators are creative in a controlled manner and observe the result. This led to the hypothesis concerning how creativity is regulated that was explored in experiment two. Experiment two in turn led to the hypothesis explored in experiment three concerning the conditions under which the benefits of social regulation of creativity are long term. The fact that each experiment yielded insights that led to a new hypothesis speaks to the value of the approach. Although further investigation is needed to establish the relevance of these results to real societies, we believe they constitute an important step forward to understand the underlying mechanisms that enable societies to balance novelty with continuity. 

\bibliographystyle{apacite} 
\bibliography{example} 
\begin{description}
\setlength{\itemsep}{0mm}

\item Adams, N. (2012). Walden Two: An anticipation of positive psychology? {\it Review of General Psychology, 16,} 1-9.  



\item Aljughaiman, A., \& Mowrer-Reynolds, E. (2005). Teachers' conceptions of creativity and creative students. \emph{Journal of Creative Behavior, 39,} 17-34.


\item Amabile, T. (1998). How to kill creativity. \emph{Harvard Business Review, Sept.- Oct.}, 77-87. 

\item Anderson,  M. L., Richardson, M. J. \& Chemero, J. (2012). Eroding the boundaries of cognition: Implications of embodiment. \emph{Topics in Cognitive Science, 4,} 717-730.



\item Baer, J. (1996). The effects of task-specific divergent-thinking training. \emph{Journal of Creative Behavior, 30,} 183-187.

\item Bandura, A. (1965). Behavioral modification through modeling procedures. In L. Krasner \& L. P. Ulmann (Eds.), \emph{Research in behavior modification: new development and implications} (pp. 310-340). New York: Rinehart and Winston.

\item Basadur, M. (1995). \emph{The power of innovation.} New York: Pitman.



\item Beghetto, R. A. (2007). Ideational code-switching: Walking the talk about supporting student creativity in the classroom. \emph{Roeper Review, 29,} 265-270.



\item Best, M. (1999). How culture can guide evolution: An inquiry into gene/meme enhancement and opposition. \emph{Adaptive Behavior, 132,} 289-293.

\item Bhattacharyya, S., \& Ohlsson, S. (2010). Social creativity as a function of agent cognition and network properties: A computer model. \emph{Social Networks, 32,} 263-278.

\item Boyd, R. \& Richerson, P. J. (1985). \emph{Culture and the evolutionary process.} Chicago: The University of Chicago Press.


\item Chan, J., \& Schunn, C. (2015). The impact of analogies on creative concept generation: Lessons from an in vivo study in engineering design. \emph{Cognitive Science, 39,} 126-155.


\item Costello, F. J., \& Keane, M. T. (2000). Efficient creativity: Constraint guided conceptual combination, \emph{Cognitive Science, 24,} 299-4349.

\item Craft, A. (2005). \emph{Creativity in schools: Tensions and dilemmas.} London: Routledge.

\item Cropley, D., Cropley, A., Kaufman, J., \& Runco, M. (2010). \emph{The dark side of creativity.} Cambridge, UK: Cambridge University Press.



\item Dantzig, S. V., Raffone, A., \& Hommel, B. (2011). Acquiring contextualized concepts: A connectionist approach. \emph{Cognitive Science, 35,} 1162-1189.

\item Dasgupta, S. (1994). \emph{Creativity in invention and design.} New York: Cambridge University Press.

\item Donald, M. (1991). \emph{Origins of the modern mind: Three stages in the evolution of culture and cognition.} Cambridge, MA: Harvard University Press.

\item Dunbar, K. (2000). How scientists think in the real world: Implications for science education. \emph{ Journal of Applied Developmental Psychology, 21,} 49-58.

\item Enquist, M., Eriksson, K. \& Ghirlanda, S. (2007). Critical social learning?: A solution to Rogers' paradox of nonadaptive culture. \emph{American Anthropologist, 109,} 727-734.



\item Feinstein, J. S. (2006). \emph{The nature of creative development.} Stanford: Stanford University Press.


\item Florida, R. (2002). \emph{The rise of the creative class.} London: Basic Books.

\item Forrest, S. \& Mitchell, M. (1993). Relative building block fitness and the building block hypothesis. In L. Whitley (Ed.), \emph{Foundations of genetic algorithms.} San Mateo: Morgan Kaufman. 


\item Gabora, L. (1995). Meme and variations: A computational model of cultural evolution. In L. Nadel \& D. Stein (Eds.), \emph{1993 lectures in complex systems.} Reading MA: Addison-Wesley.


\item Gabora, L. (2008a). EVOC: A computer model of the evolution of culture. In V. Sloutsky, B. Love \& K. McRae (Eds.), \emph{Proceedings of the 30th Annual Meeting of the Cognitive Science Society} (pp. 1466-1471). North Salt Lake, UT: Sheridan Publishing.

\item Gabora, L. (2008b). The cultural evolution of socially situated cognition. \emph{Cognitive Systems Research, 9,} 104-113. 

\item Gabora, L. (2011). Five clarifications about cultural evolution. \emph{Journal of Cognition and Culture, 11,} 61-83. 

\item Gabora, L. (2013). An evolutionary framework for culture: Selectionism versus communal exchange. \emph{Physics of Life Reviews, 10,} 117-145. 

\item Gabora, L. (2017). Honing theory: A complex systems framework for creativity. \emph{Nonlinear Dynamics, Psychology, and Life Sciences, 21,} 35-88. 

\item Gabora, L. \& Aerts, D. (2009). A mathematical model of the emergence of an integrated worldview. \emph{Journal of Mathematical Psychology, 53,} 434-451.

\item Gabora, L., Chia, W., \& Firouzi, H. (2013). A computational model of two cognitive transitions underlying cultural evolution. In M. Knauff, N. Sebanz, Michael Pauen, \& I. Wachsmuth (Eds.), \emph{ Proceedings of the 35th Annual Meeting of the Cognitive Science Society} (pp. 2344-2349). Austin TX: Cognitive Science Society. 

\item Gabora, L., \& DiPaola, S. (2012). How did humans become so creative? In M. L. Maher, K. Hammond, A. Pease, R. Pèrez, D. Ventura \& G. Wiggins (Eds.), \emph{Proceedings of the 3rd International Conference on Computational Creativity} (pp. 203-210). Palo Alto, CA: Association for the Advancement of Artificial Intelligence. 


\item Gabora, L. \& Kauffman, S. (2016). Toward an evolutionary-predictive foundation for creativity. \emph{Psychonomic Bulletin \& Review, 23}, 632-639..

\item Gabora, L., \& Kitto, K. (2013). Concept combination and the origins of complex cognition. In 
L. Swan (Ed.), \emph{Origins of mind} (pp. 361-382). Berlin: Springer.



\item Gabora, L., O'Connor, B., \& Ranjan, A. (2012). The recognizability of individual creative styles within and across domains. \emph{Psychology of Aesthetics, Creativity, and the Arts, 6,} 351-360.


\item Goldstone, R. L., \& Gureckis, T. M. (2009). Collective behavior. \emph{Topics in Cognitive Science, 1,} 412-438.


\item Guardiola, X., Diaz-Guilera, A., Perez, C., Arenas, A., Llas, M. (2002). Modeling diffusion of innovations in a social network. \emph{Physical Review E, 66,} 026121. 

\item Haldane, J. B. S. (1932/1990). \emph{The causes of evolution.} Princeton NJ: Princeton Science Library. 

\item Henrich, J., \& Boyd, R. (2002). On modeling cognition and culture: Why replicators are
not necessary for cultural evolution. \emph{Journal of Cognition and Culture, 2,} 87-112.

\item Higgs, P. (2000). The mimetic transition: a simulation study of the evolution of learning by imitation. \emph{Proceedings of the Royal Society B -- Biological Sciences, 267,} 1355-1361.

\item Hills, T., Todd, P. M., Lazer, D., Redish, A. D., Couzin, I. D., \& the Cognitive Search Research Group (2015). Exploration versus exploitation in space, mind, and society. \emph{Trends in Cognitive Science, 19,} 46-54.


\item Hinton, G., \& Nowlan, S. (1987). How learning can guide evolution. \emph{Complex Systems, 267,} 495-502.

\item Hocevar, D. (1976). Dimensionality of creativity. \emph{Psychological Reports, 39,} 869-870.

\item Holland, J. (1975). {\it Adaptation in natural and artificial systems.} Ann Arbor: University of Michigan Press.

\item Hunter, S. T., Thoroughgood, C. N., Myer, A. T., \& Ligon, G. S. (2011). Paradoxes of leading innovative endeavors: Summary, solutions, and future directions. {\it Psychology of Aesthetics, Creativity, and the Arts, 5,} 54-66.

\item Hutchins, E., \& Hazelhurst, B. (1991). Learning in the cultural process. In C. Langton, J. Taylor, D. Farmer, \& S. Rasmussen (Eds.), \emph{Artificial life II.} Redwood City: Addison-Wesley.

\item Iribarren J. L., \& Moro, E. (2011). Affinity paths and information diffusion in social networks. \emph{Social Networks 33,} 134-142. 

\item Jackson, M., Yariv, L., (2005). Diffusion on social networks. \emph{Economie Publique, 16,} 3-16.

\item Jacobs, J. (2000). \emph{The nature of economies.} New York: The Modern Library. 


\item Kameda, T., \& Nakanishi, D. (2003). Does social/cultural learning increase human adaptability? Rogers' question revisited. \emph{Evolution and Human Behavior, 24,} 242-260.


\item Kaufman, J. C. (2003). The cost of the muse: Poets die young. \emph{Death Studies, 27,} 813-822.

\item Kaufman, J. C. \& Beghetto, R. A. (2014). Creativity in schools: Renewed interest and promising new directions. In Gilman, R., Huebner, E. S., \& Furlong, M. J.  (Eds.). \emph{Handbook of positive psychology in schools} (2nd Ed.) (pp. 165-175). New York: Routledge.

\item Kharratzadeh, M., Montrey, M., Metz, A. \& Shultz, T. (2015). Resolving Rogers' paradox with specialized hybrid learners. In R. Dale, C. Jennings, P. Maglio, T. Matlock, D. Noelle, A. Warlaumont \& J. Yashimi (Eds.), \emph{Proceedings of the 37th Annual Meeting of the Cognitive Science Society} (pp. 1069-1074). Austin TX: Cognitive Science Society. 

\item Legare, C. H., Wen, N. J., Herrmann, P. A., \& Whitehouse, H. (2015). Imitative flexibility and the development of cultural learning. \emph{Cognition, 142,} 351-361. 

\item Leijnen, S., \& Gabora, L. (2009a). The artist loft effect in the clustering of creative types: A computer simulation. In N. Bryans-Kins (Ed.), \emph{Proceedings of 7th Conference on Creativity and Cognition} (pp. 389-390). New York: Association for Computing Machinery Press.

\item Leijnen, S. \& Gabora, L. (2009b). The tradeoff between degree of creativity and number of creators in a computational model of society. In B. Cooper \& V. Danos (Eds.), \emph{Proceedings of the First International Conference on Developments in Computational Models from Nature} (pp. 108-119). Sydney: Electronic Proceedings in Theoretical Computer Science.

\item Liu, B. S.C., Madhavan, R., \& Sudharshan, D., (2005). Diffunet: The impact of network structure on diffusion of innovation. \emph{European Journal of Innovation Management 8,} 240-262.

\item Ludwig, A. (1995). \emph{The price of greatness.} New York: Guilford Press.

\item Maslow, A. (1959). Creativity in self-actualizing people. In H. Anderson (Ed.), \emph{Creativity and its cultivation.} New York: McGraw-Hill.

\item May, R. (1975). \emph{The courage to create.} New York: Bantam.


\item McDonald, R., \& Siegel, D. (1986). The value of waiting to invest. \emph{The Quarterly Journal of Economics, 101}, 707-727.

\item Mesoudi, A., Whiten, A. \& Laland, K. N. (2006). Towards a unified science of cultural evolution. \emph{Behavioral and Brain Sciences, 29,} 329-383. 

\item Mithen S. 1998. {\it Creativity in human evolution and prehistory}. Abingdon-on-Thames: Routledge.



\item Mumford, M. D., \& Hunter, S. T. (2005). Innovation in organizations: A multi-level perspective on creativity. In F. J. Yammarino \& F. Dansereau (Eds.), \emph{Research in multi-level issues} (Vol. IV, pp. 11-74). Oxford, UK: Elsevier.

\item Nettle, D. (2006). The evolution of personality variation in humans and other animals. \emph{American Psychologist, 61}, 622-631.

\item Niazi, M. \& Hussain, A. (2011). Agent-based computing from multi-agent systems to agent-based models: A visual survey. \emph{Scientometrics, 89,} 479–499.

\item Nooteboom, B., W. Van Haverbeke, G. Duysters, V. Gilsing, \& A. Van den Oord (2007). Optimal cognitive distance and absorptive capacity. \emph{Research Policy, 36,} 1016-1034.


\item Plucker, J. A. (1998). Beware of simple conclusions: The case for the content generality of creativity. \emph{Creativity Research Journal, 11,} 179-182.

\item Plucker, J. A., Beghetto, R. A., \& Dow, G. T. (2004). Why is not creativity more important to educational psychologists? Potentials, pitfalls, and future directions in creativity research.\emph{Educational Psychologist, 39,} 83-96.


\item Ranjan, A. (2014).\emph{Understanding the creative process: Personal signatures and cross-domain interpretations of ideas.} (Doctoral dissertation). University of British Columbia, Canada.

\item Reiter-Palmon, R., \& Illies, J. J. (2004). Leadership and creativity: Understanding leadership from a creative problem-solving perspective. \emph{Leadership Quarterly, 15,} 55-77.

\item Rendell, L., Fogarty, L., \& Laland, K. N. (2010). Rogers' paradox recast and resolved: population structure and the evolution of social learning strategies. \emph{Evolution, 64,} 534-48.
 
 \item Robinson, K. (2001). \emph{Out of our minds: Learning to be creative.} New York: Capstone.

\item Rogers, A. R. (1988). Does biology constrain culture? \emph{American Anthropologist, 90,} 819-831.

\item Rogers, C. (1959). Toward a theory of creativity. In H. Anderson (Ed.) \emph{Creativity and its cultivation.} New York: Harper \& Row.


\item Rule, E. G., \& Irwin, D. W. (1988). Fostering intrapreneurship: The new competitive edge. \emph{The
Journal of Business Strategy 9,} 44-47.

\item Runco, M. A. (1987). The generality of creative performance in gifted and nongifted children. \emph{ Gifted Child Quarterly, 331,} 121-125.



\item Rumelhart, D.E., McClelland, J. L., and the PDP Research Group (Eds.). (1986). \emph{Parallel distributed processing.} Cambridge MA: Bradford/MIT Press.

\item Ruscio, J., Whitney, D. M., \& Amabile, T. M (1998). The fishbowl of creativity. \emph{Creativity Research Journal, 11,} 243-263. 



\item Scott, C. (1999). Teachers' biases toward creative children. \emph{Creativity Research Journal, 12,} 321-337.


\item Simonton, D. K. (20002). Creativity. In D. R. Snyder \& S. J. Lopez (Eds.) \emph{Handbook of positive psychology} (pp. 189-201). New York: Oxford University Press.


\item Snyder, H., Gregerson, M., \& Kaufman, J. (Eds.). (2012). \emph{Teaching creatively.} New York: Springer. 

\item Sosa, R., \& Connor, A. M. (2015). A computational intuition pump to examine group creativity: building on the ideas of others. \emph{Proceedings of the 2015 IASDR Conference: Interplay 2015.}

\item Spencer, G. M. (2012). Creative economies of scale: An agent-based model of creativity and agglomeration. \emph{Journal of Economic Geography, 12}(1), 247-271.


\item Stokols, D., Clitheroe, C., \& Zmuidzinaz, M. (2002). Qualities of work environments that promote perceived support for creativity. \emph{Creative Research Journal, 14}(2), 137-147.

\item Sulloway, F. (1996). \emph{Born to rebel.} New York: Pantheon.


\item Tardif, T. Z., \& Sternberg, R. J. (1988). What do we know about creativity? In R. J. Sternberg (Ed.), \emph{The nature of creativity: contemporary psychological perspectives} (pp. 429-440). Cambridge UK: Cambridge University Press.

\item Thagard, P., \& Stewart, T. C. (2011). The AHA! experience: Creativity through emergent binding in neural networks.\emph{Cognitive Science, 35}(1), 1-33. 


\item Tomasello, M., Kruger, A., \& Ratner, H. (1993). Cultural learning. \emph{Behavioral and Brain Sciences, 16}(03), 495-552. 

\item Tomlinson, R. (1980). A systems approach to planning in organizations: Developing a collaborative research study. {Cybernetics and Systems 11,} 355-367.

\item Torrance, E. (1963). \emph{ Guiding creative talent.} Englewood Cliffs, NJ: Prentice-Hall.



\item Vallet-Bellmunt, T. \& Molina-Morales, F. (2015). Be creative but not so much. Decreasing benefits of creativity in clustered firms. \emph{Entrepreneurship \& Regional Development,27,} 1-27.

\item Ward, T. B., Smith, S. M., \& Vaid, J., (Eds.). (1997). \emph{Creative thought: An investigation of conceptual structures and processes.} Washington, DC: APA Books.

\item Watson-Jones, R. E., Legare, C. H., Whitehouse, H., \& Clegg, J. M. (2014). Task-specific effects of ostracism on imitative fidelity in early childhood. \emph{Evolution \& Human Behavior, 35,} 204-210.

\item Watts, C., \& Gilbert, N. (2014). {\it Simulating innovation: Computer-based tools for rethinking innovation.} Cheltenham UK: Edward Elgar Publishing.

\item Weitzman, M. L. (1998). Recombinant growth. \emph{ Quarterly Journal of Economics, 113,} 331-360.

\item Westby, E. L., \& Dawson, V. L. (1995). Creativity: Asset of burden in the classroom? \emph{Creativity Research Journal, 8,} 1-11.

\item Wisdom, T. N., Song, X., \& Goldstone, R. L. (2013). Social learning strategies in networked groups. \emph{\it Cognitive Science, 37,} 1383-1425.

\item Wolfradt, U., \& Pretz, J. (2001). Individual differences in creativity: Personality, story writing, and hobbies. \emph{European Journal of Personality, 15,} 297-310.

\end{description}

\newpage
\section{Appendix: Training the Network}

To train the network, the activation of nodes is updated as follows. The relevant variables are:
\bigskip

\noindent $a_j$ = activation of $j$ \\
\noindent $t_j$ = $j^{th}$ component of input \\
\noindent $w_{ij}$ = weight on link from $i$ to $j$ \\
\noindent $\beta$ = 0.15  \\
\noindent $\theta$ = 0.5

\begin{equation}
a_{j} =  \frac{1}{(1 + e^{-\beta[\sum w_{ij}a_i+\theta]})}
\label{eq:activation}
\end{equation}

\indent For the movement node, we use the absolute value of $a_i$ (since negative movement is not possible; the least you can move is to not move at all). The comparison between input and output involves computing an error term, which is used to modify the pattern of connectivity in the network such that its responses become more correct. For input/output units the error term is computed as follows:

\begin{equation}
\delta_{j} =  (t_j - a_j)a_j(1 - a_j)
\label{eq:error-io}
\end{equation}

\indent For hidden units the error term is computed as follows:

\begin{equation}
\delta_{i} =  a_j(1 - a_j)\sum \delta_{j}w_{ij}
\label{eq:error-h}
\end{equation}                                              
\end{document}